\documentclass[fleqn,usenatbib,useAMS]{mnras}

\usepackage{graphicx}	% Including figure files
\usepackage{amsmath}	% Advanced maths commands
\usepackage{amssymb}	% Extra maths symbols
\usepackage{multicol}        % Multi-column entries in tables
\usepackage{bm}		% Bold maths symbols, including upright Greek
\usepackage{pdflscape}	% Landscape pages
\newcommand{\kms}{\,km\,s$^{-1}$} % kilometres per second
\defcitealias{BOTT-1976}{But76}
\defcitealias{1997MNRAS.284..599B}{Bow97}

\usepackage[T1]{fontenc}
\usepackage{ae,aecompl}

\usepackage{txfonts}

\title[Redshift determination of 3C\,66A]{Redshift determination of the BL\,Lac object 3C\,66A by the detection of its host galaxy cluster at \boldmath{$z=0.340$} \thanks{Based on observations carried out at the 8.1 meter telescope of Gemini
    North Observatory, Mauna Kea (Hawaii, USA).}}

\author[J. Torres-Zafra et al.]{Juanita
  Torres-Zafra$^{1,2}$\thanks{Contact e-mail:
    \href{mailto:jtzafra@fcaglp.unlp.edu.ar}{jtzafra@fcaglp.unlp.edu.ar}},
  Sergio A. Cellone$^{1,2}$\thanks{Current affiliation: Complejo Astron\'omico
  El Leoncito (CASLEO), CONICET - UNLP - UNC - UNSJ, Argentina}, Alberto Buzzoni$^{3}$, Ileana Andruchow$^{1,2}$ \newauthor and Jos\'e G. Portilla$^{4}$
\\
$^{1}$Instituto de Astrof\'{\i}sica de La Plata (CCT La Plata -
CONICET - UNLP), La Plata, Argentina\\ $^{2}$Facultad de Ciencias
Astron\'omicas y Geof\'{\i}sicas, Universidad Nacional de La Plata, Paseo del Bosque,
B1900FWA La Plata, Argentina\\ $^{3}$Osservatorio Astronomico di
Bologna (INAF), Via Ranzani 1, I-40127 Bologna,
Italy\\ $^{4}$Universidad Nacional de Colombia (UNal), Bogot\'a,
Colombia}
\date{Last updated 2015 May 22; in original form 2013 September 5}

\pubyear{2016}

\begin{document}
\label{firstpage}
\pagerange{\pageref{firstpage}--\pageref{lastpage}}
\maketitle

\begin{abstract}
The BL\,Lac object 3C\,66A is one of the most luminous extragalactic
sources at TeV $\gamma$-rays (VHE, i.e. $E >100$~GeV).
  Since TeV $\gamma$-ray radiation is absorbed by the extragalactic
  background light (EBL), it is crucial to know the redshift of the
  source in order to reconstruct its original spectral energy
  distribution, as well as to constrain EBL models. However, the
  optical spectrum of this BL\,Lac is almost featureless, so a direct
  measurement of $z$ is very difficult; in fact, the published
  redshift value for this source ($z=0.444$) has been strongly
  questioned. Based on EBL absorption arguments, several constraints
  to its redshift, in the range $0.096 < z < 0.5$, were proposed.
  Since these AGNs are hosted, typically, in early type galaxies that
  are members of groups or clusters, we have analysed
  spectro-photometrically the environment of 3C\,66A, with the goal of
  finding the galaxy group hosting this blazar. This study was made
  using optical images of a $5.5 \times 5.5$\,arcmin$^{2}$ field
  centred on the blazar, and spectra of 24 sources obtained
    with Gemini/GMOS-N multi-object spectroscopy.
%multi-object spectra of 24 objects
%  obtained with Gemini/GMOS-N (Gemini Multi-Object Spectrograph).
  We
  found spectroscopic evidence of two galaxy groups along the blazar's
  line of sight: one at $z\simeq 0.020$ and a second one at $z \simeq
  0.340$. The first one is consistent with a known foreground
  structure, while the second group here presented has six
  spectroscopically confirmed members. Their location along a red
  sequence in the colour-magnitude diagram allows us to identify 34
  additional candidate members of the more distant group. The blazar's
  spectrum shows broad absorption features that we identify as arising
  in the intergalactic medium, thus allowing us to tentatively set a
  redshift lower limit at $z_\mathrm{3C66A} \ga 0.33$. As a
  consequence, we propose that 3C\,66A is hosted in a
  galaxy that belongs to a cluster at $z=0.340$.

\end{abstract}


%%%%%%%%%%%%%%%%%%%%%%%%%%%%%%%%%%%%%%%%%%%%%%%%%%

%%%%%%%%%%%%%%%%% BODY OF PAPER %%%%%%%%%%%%%%%%%%

\section{INTRODUCTION}
\label{sec:intro}
Blazars are the class of active galactic nuclei (AGN) which display
the most extreme properties: strong and highly variable emission
across the entire electromagnetic spectrum, from radio to gamma-rays,
reaching in some cases the very high energy (VHE; i.e. $E >100$\,GeV)
domain \citep{1995PASP..107..803U}.
It is widely accepted that a blazar's observed radiation arises from a
relativistic jet pointing at small angles with respect to our line of
sight.  The spectral energy distribution (SED) of these sources is
thus dominated by non-thermal radiation, showing two bumps: the
low-frequency one (from radio to X-rays) is ascribed to synchrotron
emission, while the bump at high-frequencies (X-rays to $\gamma$-rays)
is likely produced by inverse-Compton scattering
\citep{1981ApJ...243..700K}.

Blazars are classified based on the specific characteristics of their
optical spectra. In this way, objects with significant emission-line
equivalent widths ($|EW| > 5$\,\AA, rest frame) are called flat-spectrum
radio quasars (FSRQ), while those without (or with weak) emission
lines are called BL\,Lac objects. The featureless continua that ---by
definition--- characterise the optical spectra of these latter, in
many cases prevent the determination of their redshifts. Nevertheless,
a precise knowledge of the distances of these objects is mandatory for
constraining models of their emission, considering also that these
sources are the dominant population of the extragalactic sky at high
energies \citep{2013IJMPD..2230025C}. In the absence of direct
(i.e. spectroscopic) redshift measurements, other indirect
procedures have been proposed to estimate, or at least constrain,
BL\,Lac's redshifts.

One of them uses the blazar's host galaxy as a standard candle.
Observations ---particularly those performed with the \textit{Hubble Space
  Telescope (HST)}--- have shown that BL\,Lac objects are, almost
always, found in elliptical galaxies; often luminous ellipticals
comparable to brightest galaxies in small clusters or groups of
galaxies \citep{2000ApJ...532..816U}.
\citet{STF05} give, for BL\,Lac host
galaxies, a mean absolute magnitude $M_R =-22.9$, with a very low dispersion
($\pm 1$\,mag), and a mean
effective radius $R_\mathrm{eff}= 10$\,kpc. \citet{SRC-2013}, on
the other hand, say that when the sample is limited to
$\gamma$-emitting BL\,Lacs, the hosts seem to be $\sim 0.4$\,mag
fainter: $M_R =-22.5$\,mag. Thus, the photometric (non)detection of the
host galaxy is used to set limits to the blazar's
distance. Alternatively, a similar approach can be made through
optical spectra, where the absence of host galaxy absorption lines
\citep[e.g.][]{LFTS2014} or the detection of foreground absorptions
\citep[e.g.][]{L2012} set lower limits to a BL\,Lac's redshift.

The study of the TeV spectra of blazars can also provide redshift limits.
 Commonly, for large values of $z$, VHE gamma-rays
are absorbed by the infrared component of the extragalactic background
 light (EBL), since its photons are annihilated via pair-production
 \citep[$\gamma_{\mathrm{VHE}} + \gamma_{\mathrm{EBL}} \rightarrow \,
   \mathrm{e}^{+} +\mathrm{e}^{-}$;][]{1967PhRv..155.1404G}. This
 causes a decrease in the observed flux and a softening of the
 measured spectrum. The optical depth ($\tau(z,E)$) can be calculated
 if a model of the EBL is available, along with the redshift ($z$) and
 the gamma-ray photons energy ($E$). In this way, $\tau$ depends both
 on the distance travelled by the gamma-ray photon and its energy
 \citep{1992ApJ...390L..49S}, therefore, the knowledge of this
 parameter can be used to estimate the intrinsic flux of the source,
 once de-absorbed. Conversely, if an EBL model and an intrinsic
 gamma-ray SED are assumed, the redshift can be estimated
 \citep{2010MNRAS.405L..76P}.

3C\,66A is a well-studied blazar; it was classified as a BL\,Lac
object by \cite{1987A&A...178...21M}, based on its significant optical
and X-ray variability, and was detected at VHE gamma-rays by VERITAS
\citep{2009ApJ...693L.104A}.  The synchrotron peak of this source is
located between $10^{15}$ and $10^{16}$\,Hz
\citep{2003A&A...407..453P}, therefore it can also be classified as an
intermediate-frequency peaked BL\,Lac object (IBL, or ISP blazar).

An accurate determination of $z$ is crucial to obtain the original photon
spectrum of the source. The VHE gamma-ray flux of distant sources,
such as 3C\,66A should the assumed $z=0.444$ be real, is expected to
be significantly suppressed. Even assuming a smaller value for $z$,
the total amount of gamma-ray flux emitted, calculated following the
prescription of \cite{2007A&A...471..439M} for 3C 66A, must be huge
and with a very
hard spectrum \citep{2009ApJ...692L..29A}. Hence, verification of the
actual redshift of the source is of importance both for AGN
astrophysics and cosmology, therefore, several efforts have been made
to determine this physical parameter.

The published redshift of this blazar was determined to be $z=0.444$
from optical spectroscopy by \cite{1978bllo.conf..176M}, based on the
detection of one single line; however, the authors stated that they
were not sure of the reality of this emission feature, and warned that
the redshift value is not reliable. \cite{1993ApJS...84..109L} claimed
a confirmation of the $z= 0.444$ value, based on data from the
\emph{International Ultraviolet Explorer} (IUE). Only one emission
line was marginally detected and systematic errors affecting such
detections could not be ruled out completely. This $z$ value was
questioned by \cite{2005ApJ...629..108B}, who encouraged further
spectroscopic observations of this source. Another observation of
3C\,66A at a different spectral range was reported by
\cite{2008A&A...477..513F}, but no spectral feature was found; a lower
limit for its redshift was thus set at $z \ge 0.096$. On the other
hand \citet{1996ApJS..103..109W} reported a marginal photometric
detection of 3C\,66A's host galaxy setting a lower limit to its
redshift at $z > 0.321$. 

More recently, by assuming that the EBL-corrected TeV spectrum is not harder
than the Fermi--LAT spectrum, \cite{2010MNRAS.405L..76P} suggested an
upper-limit $z < 0.34 \pm 0.05$ for the redshift of 3C\,66A, with its
most likely value being $z= 0.213 \pm 0.05$. Also based on EBL
absorption of 3C\,66A's VHE spectrum, \citet{YW-2010} set $z < 0.58$,
while \citet{YFD10} obtained $0.15 \le z \le 0.31$, with a best-fit
value $z=0.21$.
Finally, \cite{2013ApJ...766...35F} reported a firm lower limit of
3C\,66A's redshift at $z\ge 0.3347$, while a 99.9\% confidence
upper-limit was set at $z<0.41$. These results are based on
far-ultraviolet spectra (1132\,\AA--1800\,\AA) obtained with the
Cosmic Origins Spectrograph (COS) on-board HST, where intergalactic
medium absorption features were detected on a smooth continuum.

An independent approach to find out the redshift of a BL\,Lac with
featureless optical spectrum, is to search its (projected)
neighbourhood for a group or cluster of galaxies, to which the blazar
might be associated. Although several works have undertaken the study
of the immediate environment of BL\,Lacs \citep[and references
  therein]{2016MNRAS.455..618F}, the use of the group/cluster redshift
as a proxy for the blazar's redshift has been scarcely used
\citep{PFT-1995, 2015A&A...574A.101M, RMDP-2016} \citep[see also][for
  preliminary results from the present project]{TCA-2013}.

  In particular, the galaxy environment of 3C\,66A was tested by
  \citet[hereafter But76]{BOTT-1976}, who found 15 probable members of
  a cluster surrounding 3C\,66A, from which they estimated a
  photometric redshift $z \approx 0.37^{+0.07}_{-0.03}$. Later, and
  using the luminosity function of field galaxies, \citet{WESY-1993}
  suggested that the blazar is associated to a poor cluster with a
  loosely constrained redshift $0.3 \le z \le 0.5$. \citet[hereafter
    Bow97]{1997MNRAS.284..599B}, in turn, provided redshifts between
  $z=0.0198$ and $z=0.0675$ for 7 galaxies within a 16.7\,arcmin$^{2}$
  projected area centred on 3C\,66A; however, these objects are
  relatively bright, and none of them is included among the cluster
  member candidates cited above.

Considering the controversy about the redshift of 3C\,66A, we have
studied its environment using deep, two-colour photometry of $300$
objects within a field centred on the blazar, and spectra of a
sub-sample of 24 objects (down to $g'=22.6$\,mag, and including 3C\,66A
itself), obtained with the Gemini North Telescope and the Gemini
Multi-Object Spectrograph (GMOS). Our goal is to look for both
photometric and spectroscopic signatures of a group or cluster of
galaxies hosting the blazar, thus providing an independent and
reliable determination of its redshift.

This paper is structured as follows: in Section~\ref{sec:obs} the
photometric and spectroscopic data are
described. Section~\ref{sec:resul} presents data analysis and results,
and in Section~\ref{sec:syc} the work is summarised and our
conclusions are presented.
In this paper we assume a concordance cosmology with
$H_{0}=69.6$\,\kms\,Mpc$^{-1}$, $\Omega _{\mathrm{m}}=0.286$ and
$\Omega _{\Lambda}=0.714$.
\section{OBSERVATIONS AND DATA ANALYSIS}
\label{sec:obs}
Observations were made at the 8.1\,m Gemini North Telescope, Mauna
Kea, Hawaii (USA), in August 2009, using the GMOS (Gemini Multi-Object
Spectrograph) instrument in both image and MOS modes (program
GN-2009B-Q-2; PI: I. Andruchow).  The details on these data and their
reduction process are given below.

\subsection{Images} 
\label{sec:obs.ima}
Observations included images in the $g'$ (475\,nm) and $i'$ (780\,nm) bands,
covering a field of $5.5 \times 5.5$\,arcmin$^{2}$ centred on the
blazar. Seeing $FWHM$ was about 0.7\,arcsec, and we chose to use no
binning (thus giving a scale of 0.0727 arcsec pix$^{-1}$), to avoid
saturation of the AGN, while getting sufficient S/N on the field
galaxies. For the same purpose, we took several short-exposure frames
through each filter: $11 \times 120$\,s and $13 \times 25$\,s in $g'$
and $i'$, respectively.

Images were bias-subtracted and flat-fielded with dome flats. Overall,
data reduction was performed with tasks from the \textsc{gemini.gmos}
package included in \textsc{iraf}%
\footnote{IRAF is distributed by the National Optical Astronomy
  Observatories, which are operated by the Association of Universities
  for Research in Astronomy, Inc., under cooperative agreement with
  the National Science Foundation.} % software.

\begin{figure}
\includegraphics[width=1\columnwidth]{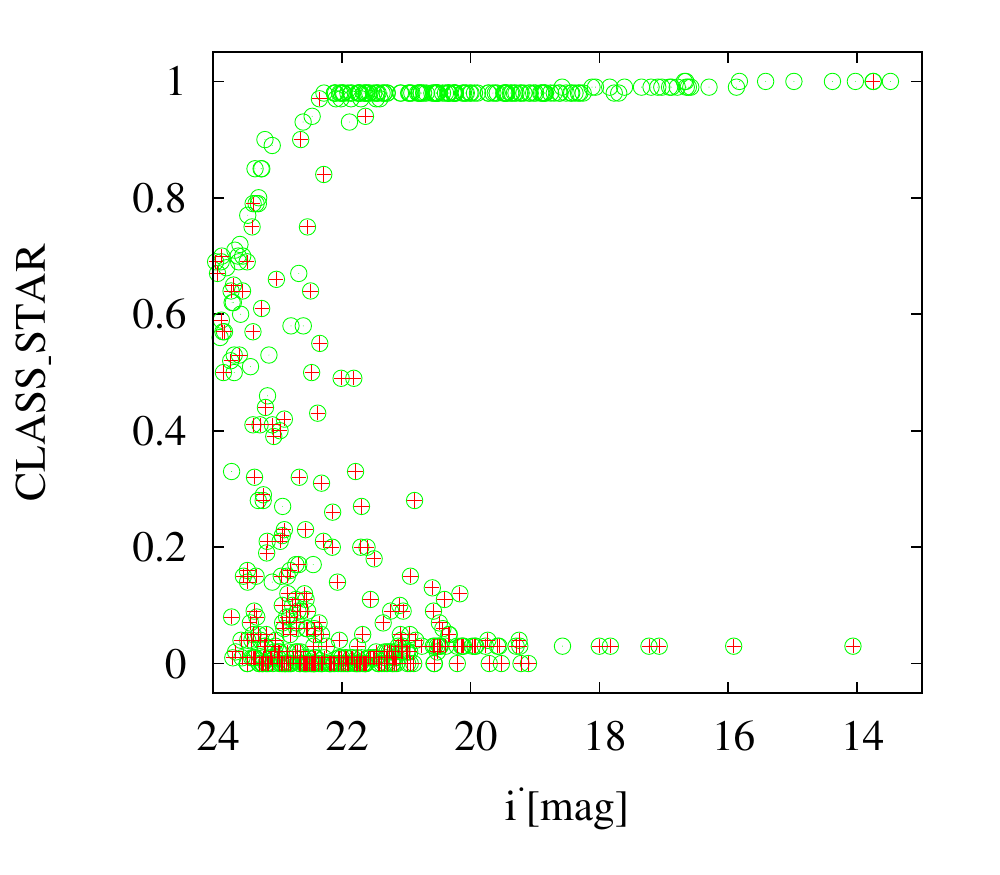}
\caption{\textsc{SExtractor} classifier (\textit{CLASS\_STAR}) as a
  function of the apparent magnitude $i'$ of all the sources detected
  in the field of 3C\,66A (green circles). \textit{CLASS\_STAR} $=1$:
  point sources; \textit{CLASS\_STAR} $=0$: extended sources.
  Galaxies detected in the field having $FWHM>8$\,pixels in the $i'$
  band (plus the blazar, the $i'\approx 14$ object with
  \textit{CLASS\_STAR} $\approx 1$) are shown with red crosses.}
\label{fig:stell}
\end{figure}

We performed photometry on the images using \textsc{SExtractor}
\citep{sex}. This software allows to identify and classify point and
extended sources, building catalogues with several photometric
parameters. The \textit{MAG\_AUTO} parameter was used to measure
total magnitudes of all the objects detected in the field, with an average
error of $\pm 0.005$\,mag for the brightest sources ($g'<21$) and $\pm
0.026$\,mag for the faintest sources ($g'>23$).  The
\textit{CLASS\_STAR} parameter, in turn, allowed us to identify
the galaxies. This classification was further verified from the PSF
(Point Spread Function) of the sources, discarding those showing a
brightness profile with $FWHM \leq 8$\,pixels in the $i'$
band. In this way we could discriminate, with greater certainty,
extended sources presenting intermediate classifications with %the
\textit{CLASS\_STAR} %parameter of \textsc{SExtractor}
 (see Fig.~\ref{fig:stell}). 
A visual inspection of the identified extended
objects was performed, discarding those which were found to be
saturated point sources, CCD blemishes, etc.

%\begin{figure}
%\centering
%\includegraphics[width=\columnwidth]{CMD_galax_star.pdf}
%\caption{Colour-magnitude diagram of all sources detected in the field
%  of 3C\,66A (green circles). Objects identified as galaxies based on
%  the criterion $FWHM > 8$\,pixels are shown with red crosses.}
%\label{fig:CMD}
%\end{figure}

Magnitudes were corrected for Galactic extinction with the values
from \citet{SF11}, as 
given in \textit{NASA Extragalactic
  Database (NED)}%
\footnote{\url{https://ned.ipac.caltech.edu/}}. %
We compared our photometry with several field stars in common with
\citet{2001AJ....122.2055G}, obtaining fairly good agreement. The apparent
magnitudes for 3C\,66A were then measured to be $g'=14.32$
and $i'=13.75$\,mag. In Fig.~\ref{fig:stell}, we plot the
classification parameter \textit{CLASS\_STAR} as a function of the
$i'$ magnitude for all sources detected in the field of 3C\,66A. 
On
the same diagram, we plotted extended objects identified from the
$FWHM > 8$\,pixels criterion, which were selected for this research.

The objects selected within the field for further photometric analysis were
those brighter than 24.5\,mag in the \textit{g$'$} band %(see Fig.~\ref{fig:CMD})
.

\begin{figure*}
\includegraphics[width=0.75\textwidth]{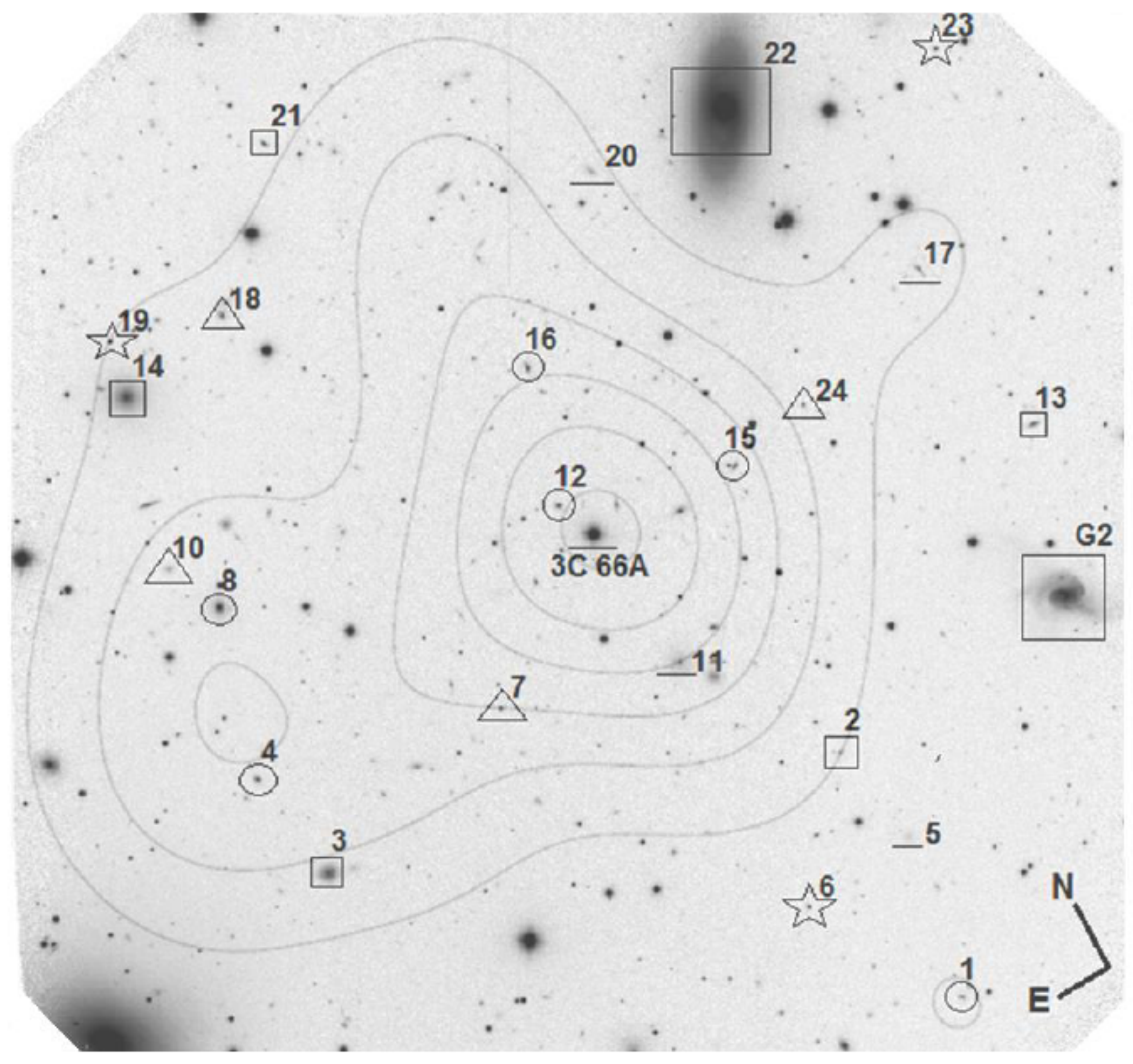}
\caption{GMOS $i'$-band image of the field centred on
  3C\,66A. Spectroscopically observed objects are labelled following their
  slit number and marked according to their redshift (see Table~\ref{tab:a}),
  with respect to the detected $z=0.340$ group (see Section~\ref{sec:idgalg}):
  triangles ($z > 0.340$) , circles ($z \simeq 0.340$), squares ($z <
  0.340$), star shape (Galactic stars), and straight lines underneath
  (objects with unknown redshift). The large boxes mark galaxies G1
  (slit \#22) and G2 in \citet{1997MNRAS.284..599B}. Contours show the 
  projected density of candidate members of the $z=0.340$ group (see
  text). The frame is 5.5 arcmin on a side, corresponding to 1.61
    Mpc at $z=0.340$.}
\label{fig:image}
\end{figure*}

\subsection{Spectra}

Spectroscopic data were acquired in the multi-object (MOS) mode,
making use of two diffraction gratings: $B600_{-}G5307$ and
$R400_{-}G5325$. Two multi-slit masks (one for each grating) were
created from a co-added image provided by Gemini from our photometric
images (see Section~\ref{sec:obs.ima}). The final mask for the
B600$_{-}$G5307 grating consisted of 23 slits with dimensions of
1.5\,arcsec by 8\,arcsec, whereas for the $R400_{-}G5325$ grating
it consisted of 11 slits with dimensions of 1\,arcsec by
8\,arcsec. Slits were placed prioritising faint non-stellar
  objects in the usable field. Six exposures of 1800\,s each were
obtained with the $B600_{-}G5307$ grating (with a binning of $2\times
2$) in three different central wavelengths: 490\,nm, 500\,nm and
510\,nm, to remove the gaps between the CCDs. The wavelength range
covered was 3500\,\AA--6500\,\AA\,\, with a spectral resolution of
0.54\,nm. The $B600_{-}G5307$ mask allowed us to extract the
spectra of 23 objects.

The $R400_{-}G5325$ grating was used in the MOS+N\&S mode, covering a
wavelength range of 7000\,\AA--10\,000\,\AA\ with a spectral
resolution of 0.79\,nm. Eight exposures of 500\,s each were obtained
(with a binning of $2\times 2$) in three different central
wavelengths: 790\,nm, 800\,nm and 810\,nm. The \textit{Nod\&Shuffle}
(N\&S) mode was used to optimally subtract the sky \citep{GB-2001},
which is a particularly important issue in the redder
range. Unfortunately, an error during the observations design led to
partial overlapping of the resulting spectra, so we were only able to
extract the spectra of 4 objects in this mask (of these, 3 were also
observed through the $B600_{-}G5307$ mask).
The objects with spectroscopic data in the field (with at least one
grating) are shown in Fig.~\ref{fig:image}.

The spectroscopic data reduction %, similarly as it was done for the images,
 was also performed with standard \textsc{iraf} procedures within
the \textsc{gemini} package, namely: \textsc{gbias}, \textsc{gsflat},
\textsc{gsreduce}, \textsc{gsmosaic}, and \textsc{gscut}. Pixel to
wavelength transformation was established using \textsc{gstransform},
from CuAr lamps spectra obtained as daytime calibrations. Finally, the
individual spectra were extracted with \textsc{gsextract}. As redshift
determination was our main goal, no flux calibration was %eventually
performed on the extracted spectra.

This procedure was followed on data obtained in both observing modes;
however, additional tasks (\textsc{gnsskysub} and \textsc{gnscombine}) were
necessary for the subtraction of the sky and the combination of spectra taken
in the N\&S mode.

Cosmic rays cleaning was performed following the routine given in
the \textsc{gscrspec.cl} script, available at the Gemini
Observatory website\footnote{\url{http://www.gemini.edu/sciops/data/software/gscrspec.cl}}.
This script uses the \emph{Laplacian Cosmic Ray Identification}
method developed by Pieter G. van Dokkum in the \textsc{lacos\_spec.cl}
script\footnote{\url{http://www.astro.yale.edu/dokkum/lacosmic/lacos\_spec.cl}}.
We verified that this procedure is more
efficient than the \textsc{iraf} task \textsc{gscrrej} to clean cosmic rays.

We then used the \textsc{fxcor} task to determine the redshifts of selected
objects in the field.  This task allows to calculate the radial velocity
through Fourier cross correlation between the spectrum of the object under
analysis and a reference (template) spectrum \citep{TD79}. Both spectra are
continuum subtracted and Fourier filtered before doing the correlation,
while dispersions are equalised by rebinning to the smallest dispersion.
For this work we used two reference spectra: that of NGC\,4449
  \citep{K92}, an irregular galaxy that has well-defined emissions, and the
spectrum of NGC\,4387 \citep{GZS04}, an elliptical
galaxy showing strong absorptions. Both templates were downloaded from
\textit{NED}. For those spectra where it was possible to establish a
tentative redshift value, typical emission lines were identified, such as
H\,$\gamma$, [\ion{O}{ii}] (3727\AA), H\,$\beta$, [\ion{O}{iii}] (4959\AA),
[\ion{O}{iii}] (5007\AA), and/or absorptions, like \ion{Ca}{ii} (H+K),
Ca+Fe, \ion{Na}{i}, H\,$\alpha$. Finally, the redshift adopted for each
object was computed from Gaussian fits to two or more such features in its
spectrum.
Three objects (slits \#6, \#19, and \#23) turned out to be Galactic stars%
\footnote{The spectrum of object \#6 looks like that of a red dwarf star, possibly
  blended with a hotter component. Given its magnitude ($g'\simeq 21$\,mag),
  it should then be a very distant ($\approx 30$\,kpc) halo star.} (we give
their radial velocities in Table~\ref{tab:a}), while no
reliable redshift value could be obtained for other five objects (slits \#5,
\#11, \#13, \#17 and \#20), due to the low S/N ratio of their spectra (we will
return to objects \#5 and \#11 in Section~\ref{sec:CRS}).  Tentative redshifts for
two of them (slits \#17 and \#20) were measured through just one emission
line each, which we assumed to be [\ion{O}{ii}] (3727\AA) and H\,$\delta$,
respectively (based on their colours and spiral morphology). So, their tentative redshift values would be $z_{\#17} \simeq
0.476$ and $z_{\#20} \simeq 0.479$. No clear emission or absorption lines
could be identified in the blazar's spectrum, besides telluric lines and
diffuse interstellar bands (DIB), so no definite redshift value could be
established in this way for 3C\,66A (see, however,
Section~\ref{sec:spec3C66A} for our analysis on probable foreground
absorptions on the blazar's spectrum). Redshifts were thus established for 15
(plus two just tentative) out of the 24 selected objects. Our results are
shown in Table~\ref{tab:a}, where we also include the only galaxy with a
previously published redshift within our GMOS
field (G2 in
\citetalias{1997MNRAS.284..599B}).

\begin{table*}
  \caption{Spectro-photometric results for selected objects in the
    field of 3C\,66A: col.\ 1: slit number; col.\ 2: identification
    label (ID), either from this paper or from other works; cols.\ 3
    and 4: RA and Dec (J2000); cols.\ 5
    and 6: $g'$ and $i'$ total integrated magnitude;
    col.\ 7: redshift ($z$), or radial velocity (for Galactic stars) ; col.\ 8: object type.}
 \label{tab:a}
 \begin{tabular}{rlcccccc}
  \hline
 Slit &  ID & RA$_\mathrm{J2000}$ & Dec$_\mathrm{J2000}$ & $g'$ & $i'$ & $z$
 (or $v_\mathrm{r}$) &  Object type \\
  &  & (hh:mm:ss) & ($^\circ$:$'$:$''$)  & (mag) & (mag) &  &   \\
  \hline
 1  & 3C66A\_01       & 02:22:35.3 & 42:59:05 &  22.275 & 20.518 &$0.3402\pm 0.0021$    & Elliptical            \\
 2  & 3C66A\_02       & 02:22:35.7 & 43:00:32 &  21.849 & 21.309 &$0.0517\pm 0.0003$    & Spiral (strong bulge) \\
 3  & 3C66A\_03       & 02:22:50.7 & 43:01:03 &  18.859 & 17.830 &$0.1521\pm 0.0007$    & Spiral (Sa?)          \\
 4  & 3C66A\_04       & 02:22:51.4 & 43:01:39 &  21.354 & 19.581 &$0.3390\pm 0.0004$    & Spiral                \\
 5  & 3C66A\_05       & 02:22:34.9 & 42:59:58 &  22.012 & 20.974 & \ldots              & Dwarf elliptical\\
 6  & 3C66A\_06       & 02:22:38.3 & 42:59:51 &  21.992 & 19.231 & ($109.9 \pm 71.4$ km\,s$^{-1}$) & Galactic star         \\
 7  & 3C66A\_07       & 02:22:44.1 & 43:01:28 &  21.827 & 20.575 &$0.5355\pm 0.0008$    & Spiral                \\
 8  & 3C66A\_08       & 02:22:50.4 & 43:02:34 &  20.060 & 18.001 &$0.3402\pm 0.0006$    & Elliptical            \\
 9  & \textbf{3C\,66A}& 02:22:39.6 & 43:02:08 &  14.323 & 13.755 &  \ldots    & Blazar       \\
10  & 3C66A\_10       & 02:22:51.3 & 43:02:52 &  21.489 & 20.559 &$0.4550\pm 0.0006$    & Irregular             \\
11  & 3C66A\_11       & 02:22:38.8 & 43:01:18 &  20.065 & 19.100 & \ldots   & Dwarf elliptical      \\
12  & 3C66A\_12       & 02:22:40.2 & 43:02:20 &  21.828 & 19.730 &$0.3408\pm 0.0010$   & Elliptical             \\
13  & 3C66A\_13       & 02:22:26.7 & 43:01:42 &  21.138 & 19.768 & \ldots  & Spiral or S0         \\
14  & 3C66A\_14$^{a}$  & 02:22:50.4 & 43:03:47 &  18.124 & 17.072 &$0.0200\pm 0.0006$   & Dwarf elliptical       \\
15  & 3C66A\_15       & 02:22:35.1 & 43:02:09 &  22.116 & 19.933 &$0.3401\pm 0.0005$    & Elliptical            \\
16  & 3C66A\_16       & 02:22:39.4 & 43:03:04 &  21.383 & 19.247 &$0.3398\pm 0.0010$    & Elliptical            \\
17  & 3C66A\_17$^{*}$  & 02:22:27.9 & 43:02:42 &  22.186 & 20.208 & $0.476\pm 0.0008$    &  Spiral               \\
18  & 3C66A\_18       & 02:22:46.9 & 43:03:59 &  21.529 & 19.559 &$0.4930\pm 0.002$     & Early-type Spiral     \\
19  & 3C66A\_19       & 02:22:50.2 & 43:04:06 &  22.053 & 20.107 & ($-83.0 \pm 9.5$ km\,s$^{-1}$) & Galactic star         \\
20  & 3C66A\_20$^{*}$  & 02:22:35.4 & 43:03:53 &  21.579 & 20.107 & $0.479\pm 0.0007$    & Late-type spiral        \\
21  & 3C66A\_21       & 02:22:43.8 & 43:04:45 &  20.963 & 20.107 &$0.1755\pm 0.0003$    & Spiral (Sc?)          \\
22  & 3C66A\_22$^{b}$  & 02:22:31.1 & 43:03:54 &  15.342 & 14.067 &$0.0201\pm 0.0007$    & Spiral (Sa)          \\
23  & 3C66A\_23       & 02:22:24.8 & 43:03:44 &  21.701 & 20.258 & ($31.0 \pm 95.3$ km\,s$^{-1}$) & Galactic star          \\
24  & 3C66A\_24       & 02:22:32.6 & 43:02:18 &  21.710 & 20.478 &$0.4275\pm 0.0007$   & Spiral                 \\
- -  & G2$^{c}$        & 02:19:18.3 & 42:47:09 &  17.090 & 15.917 & 0.0667              & Spiral                \\
\hline
\multicolumn{8}{l}{$^a$ Galaxy G3 in \citetalias{1997MNRAS.284..599B} (no redshift was reported).}\\
\multicolumn{8}{l}{$^b$ UGC\,1832 $\equiv$ galaxy G1 in \citetalias{1997MNRAS.284..599B}; the published value $z=0.0198$ agrees with our result.}\\
\multicolumn{8}{l}{$^c$ Redshift from \citetalias{1997MNRAS.284..599B}.}\\
\multicolumn{8}{l}{$^*$ Galaxies whose tentative redshifts were measured through just one emission
line.}\\
\end{tabular}
\end{table*}

\section{RESULTS}
\label{sec:resul}

In order to determine or ---at least--- to constrain the redshift of 3C\,66A, we perform
an individual spectroscopic analysis of the source itself, as well as a
spectro-photometric analysis of its close environment,
aiming to identify the host
galaxy group of this BL\,Lac object. % and set its redshift.
Considerations about the blazar's host galaxy are also presented.

\subsection{Identification of galaxy groups}
\label{sec:idgalg}

\begin{figure}
\centering \includegraphics[width=\columnwidth]{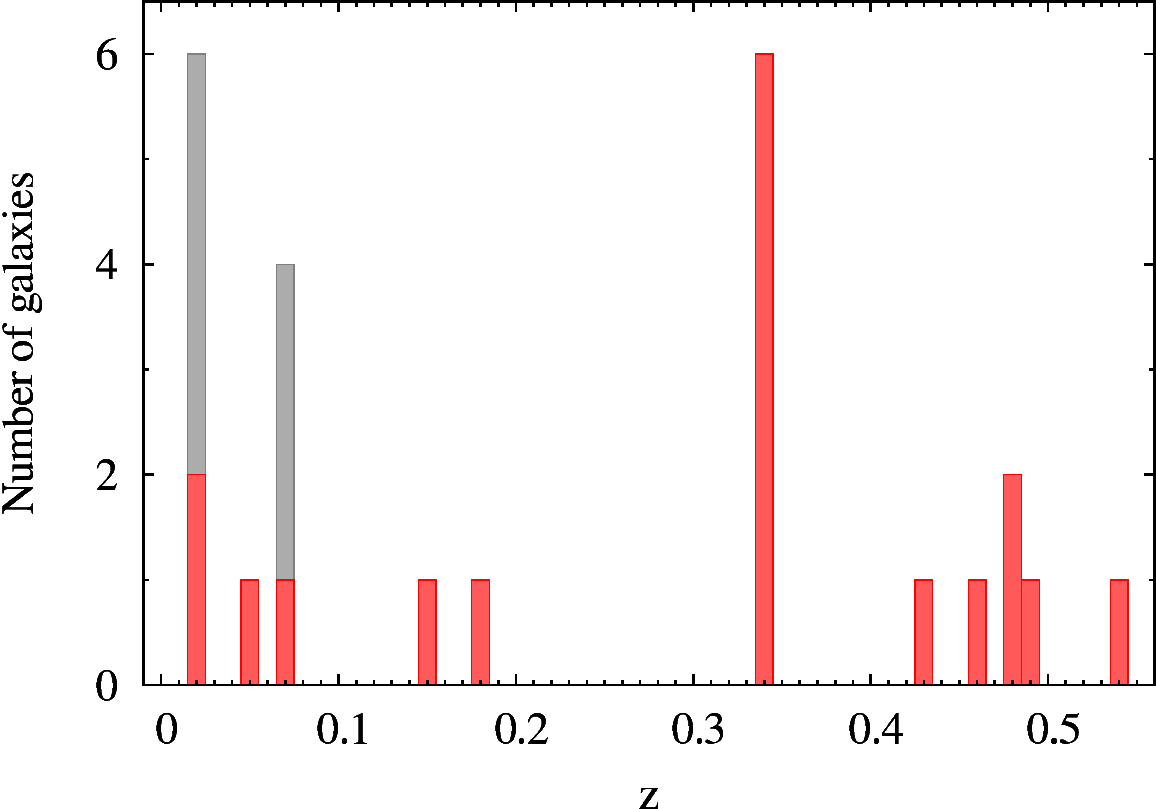}
\caption{Redshift distribution of observed sources in a field of radius
  7\,arcmin centred on 3C\,66A. Each bar represents the total number of
  objects with spectroscopic redshifts, both within (red) and outside
  (grey) the field studied in this paper.  The histogram shows three galaxy
  groupings, at $z \approx 0.34$, $z \approx 0.067$ and at $z \approx 0.02$
  in the close (projected) environment of the blazar.}
\label{fig:hz}
\end{figure}

Based on the spectro-photometric results obtained (Table~\ref{tab:a}), we
plot in Fig.~\ref{fig:hz} the distribution of the 18 spectroscopic
redshifts measured within the area of our GMOS frame (17 from the present
study plus one from the literature).  We include 8 additional galaxies with
published redshifts (from NED) lying within 7 arcmin from 3C\,66A, but out
of our $5.5 \times 5.5$ arcmin field.

At least 3 concentrations of galaxies are evident in redshift space:
one at $\langle z \rangle =0.020$, a second at $\langle z \rangle
=0.067$, and a third one at $\langle z \rangle =0.340$.  The first (6
members) is close to the redshift of the cluster Abell\,347
\citep[$z=0.0184$;][]{A1958}. In particular, the sight-line to 3C\,66A
goes through the poor cluster WBL\,069, associated to Abell\,347, and
composed by the galaxies UGC\,1832, UGC\,1837, UGC\,1841
(a.k.a. 3C\,66B), and its compact dwarf neighbour V\,Zw\,230,
with a mean redshift $\langle z \rangle=0.02088$ \citep{WBB1999}.
UGC\,1832 is galaxy G1 in \citetalias{1997MNRAS.284..599B}; we obtain
for it $z=0.0201$ (slit \#22), in agreement with the published
redshift.  We obtain $z=0.0200$ for the galaxy in slit \#14;
\citetalias{1997MNRAS.284..599B} list this galaxy as G3, with no
redshift value due to the low S/N in their spectrum. We thus confirm it as
a member of the same foreground group.

A second group in redshift space comprises 4 galaxies within 5\,arcmin
from 3C\,66A with redshifts between $z= 0.0674 \to 0.0677$ according
to NED. These include galaxies G2, G4\,\footnote{This galaxy seems to
  be misidentified in NED; its coordinates in the database point to an
  object that looks like a Galactic star.}, and G6 in
\citetalias{1997MNRAS.284..599B}, along with the radio-source 3C\,66,
although only the first one lies within the GMOS field.

\begin{figure*}
\centering
\includegraphics[width=8.8cm,height=7cm]{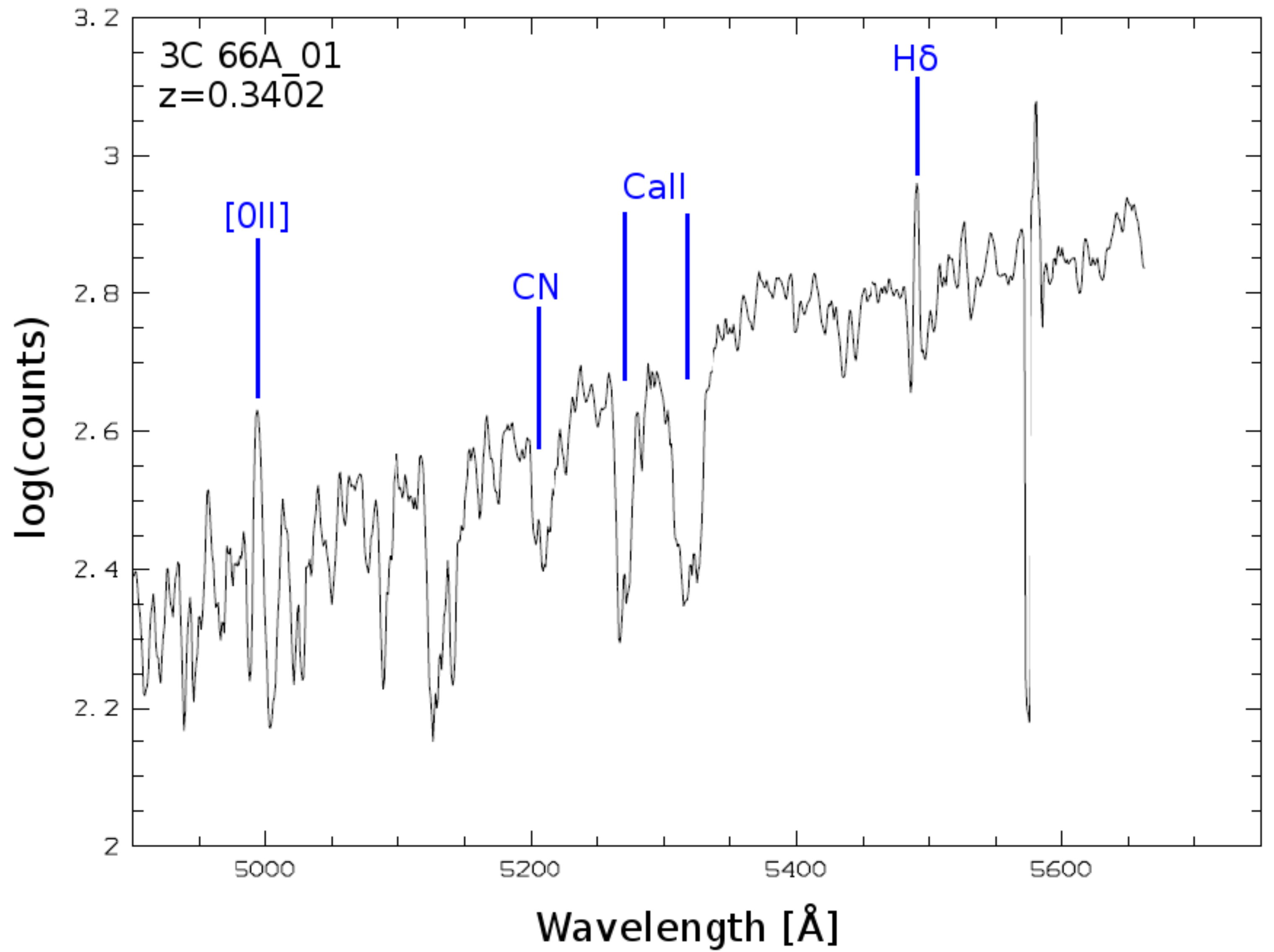}
\includegraphics[width=8.8cm,height=7cm]{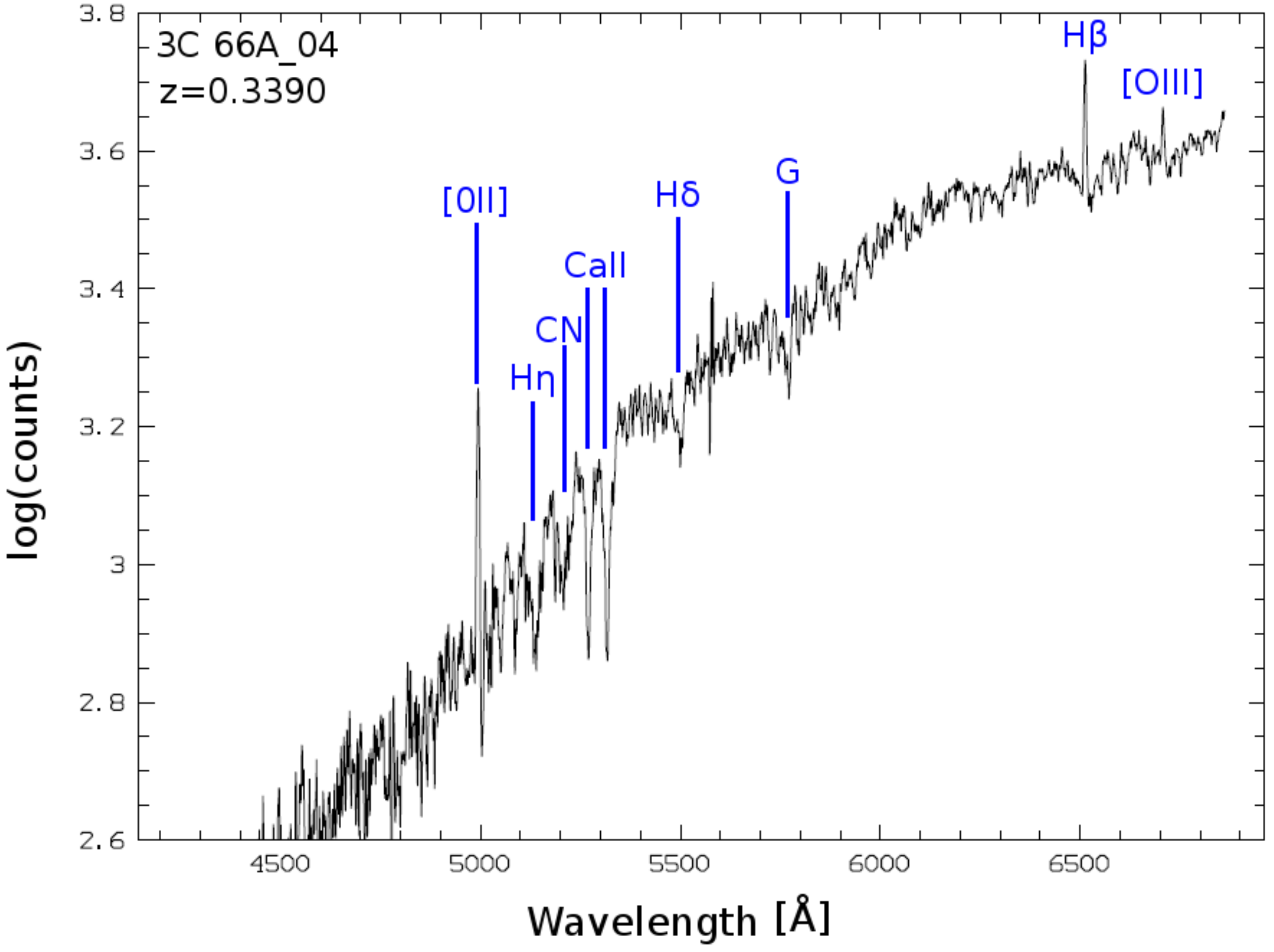}
\includegraphics[width=8.8cm,height=7cm]{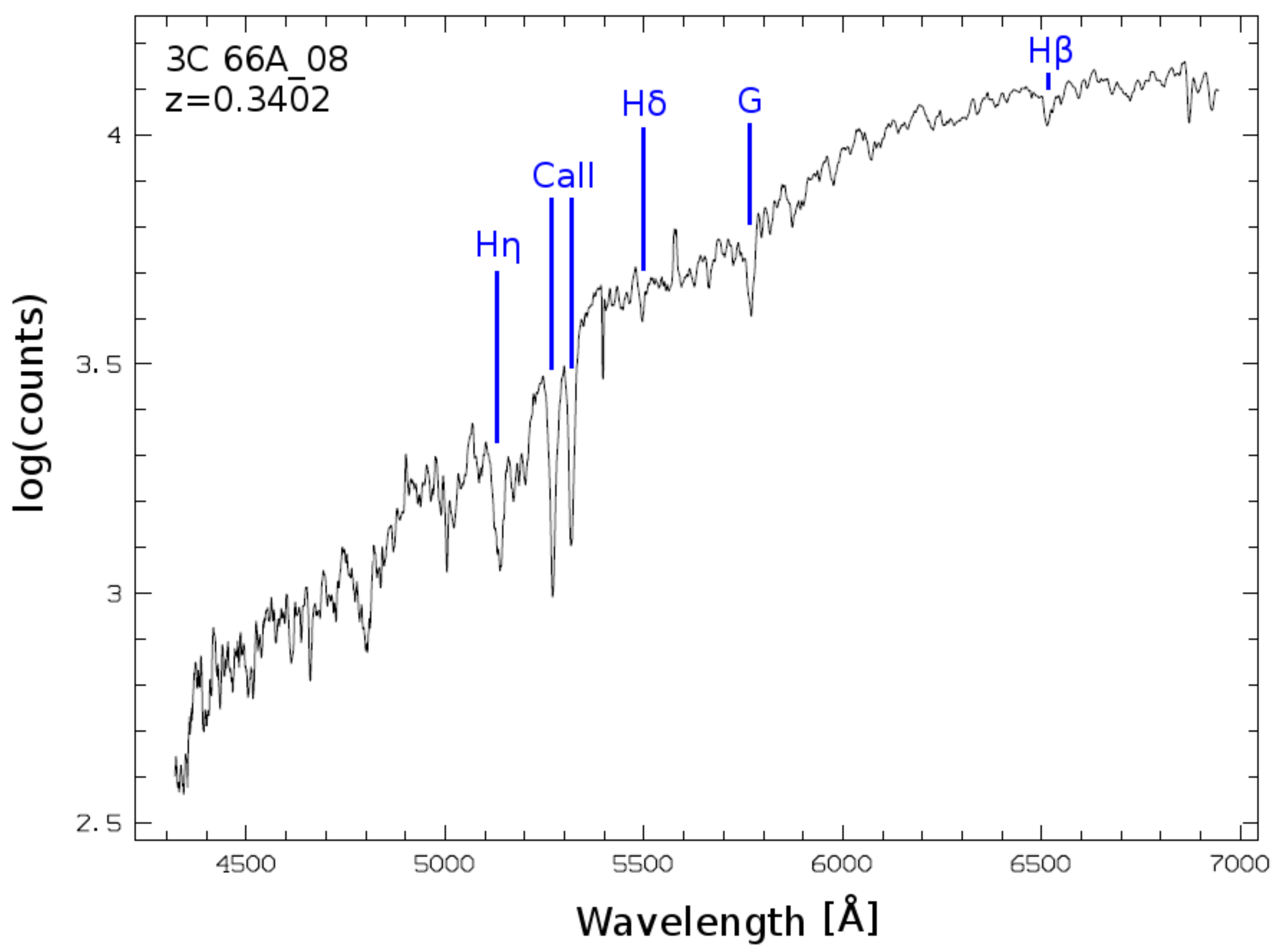}
\includegraphics[width=8.8cm,height=7cm]{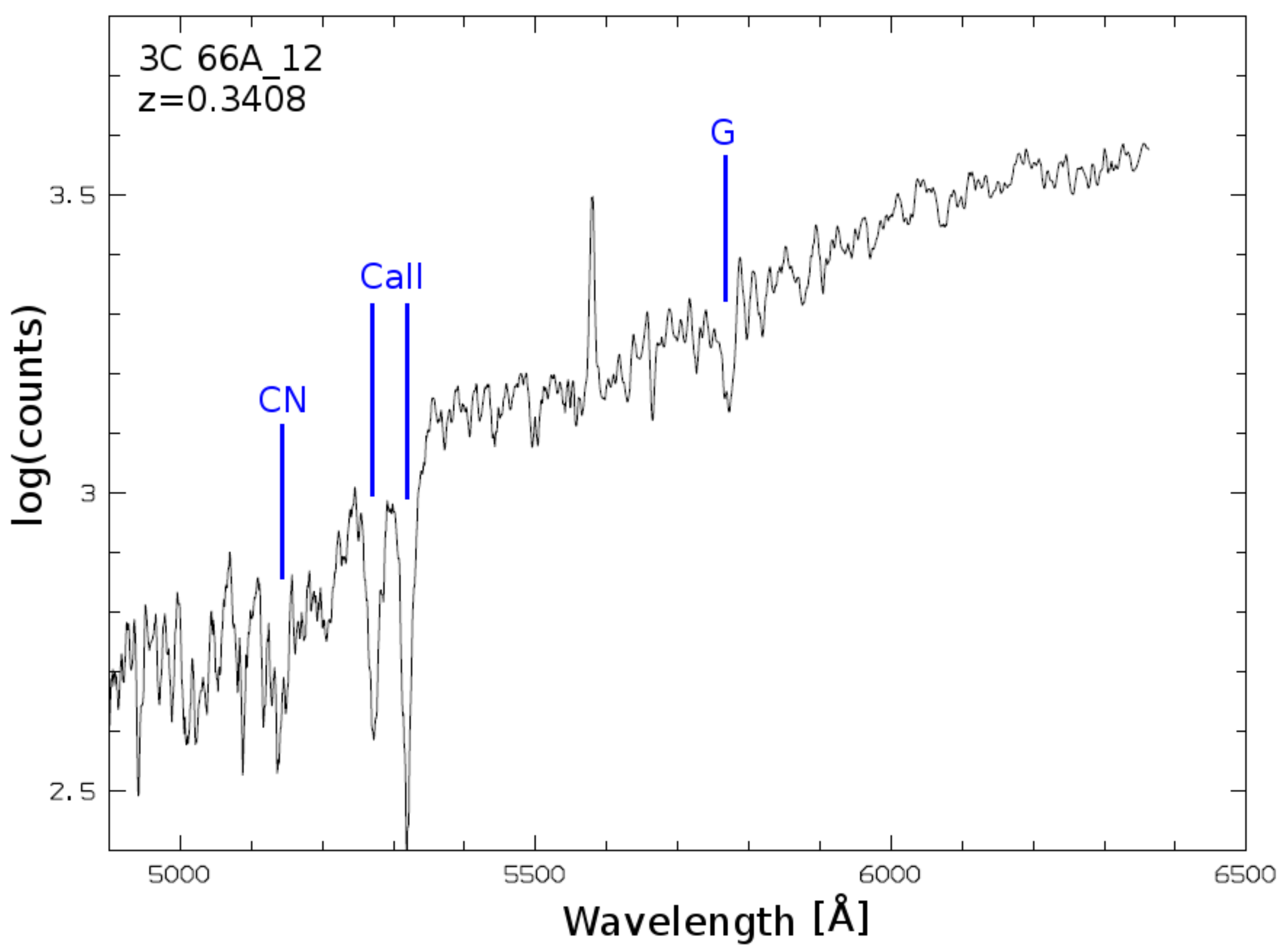}
\includegraphics[width=8.8cm,height=7cm]{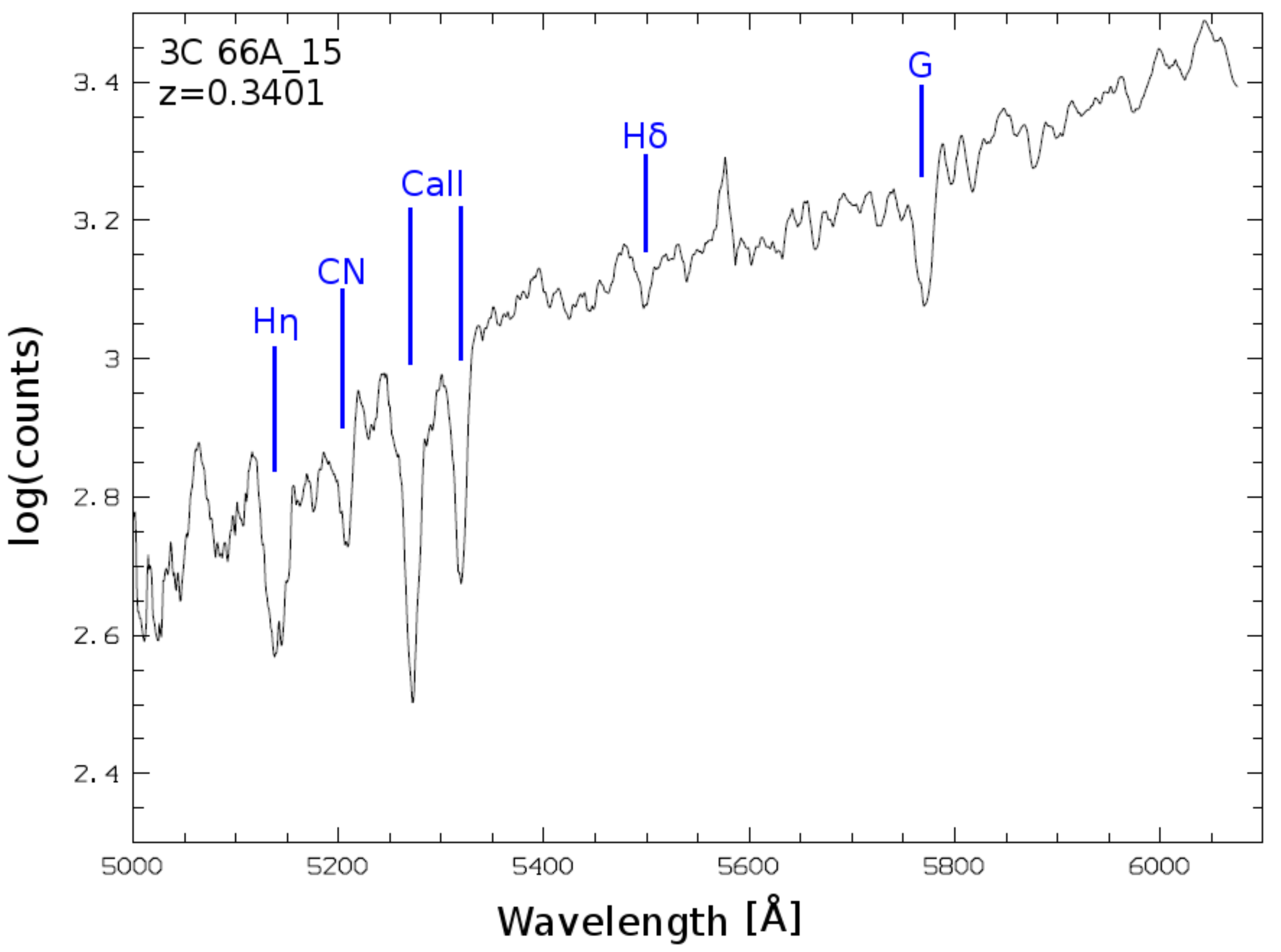}
\includegraphics[width=8.8cm,height=7cm]{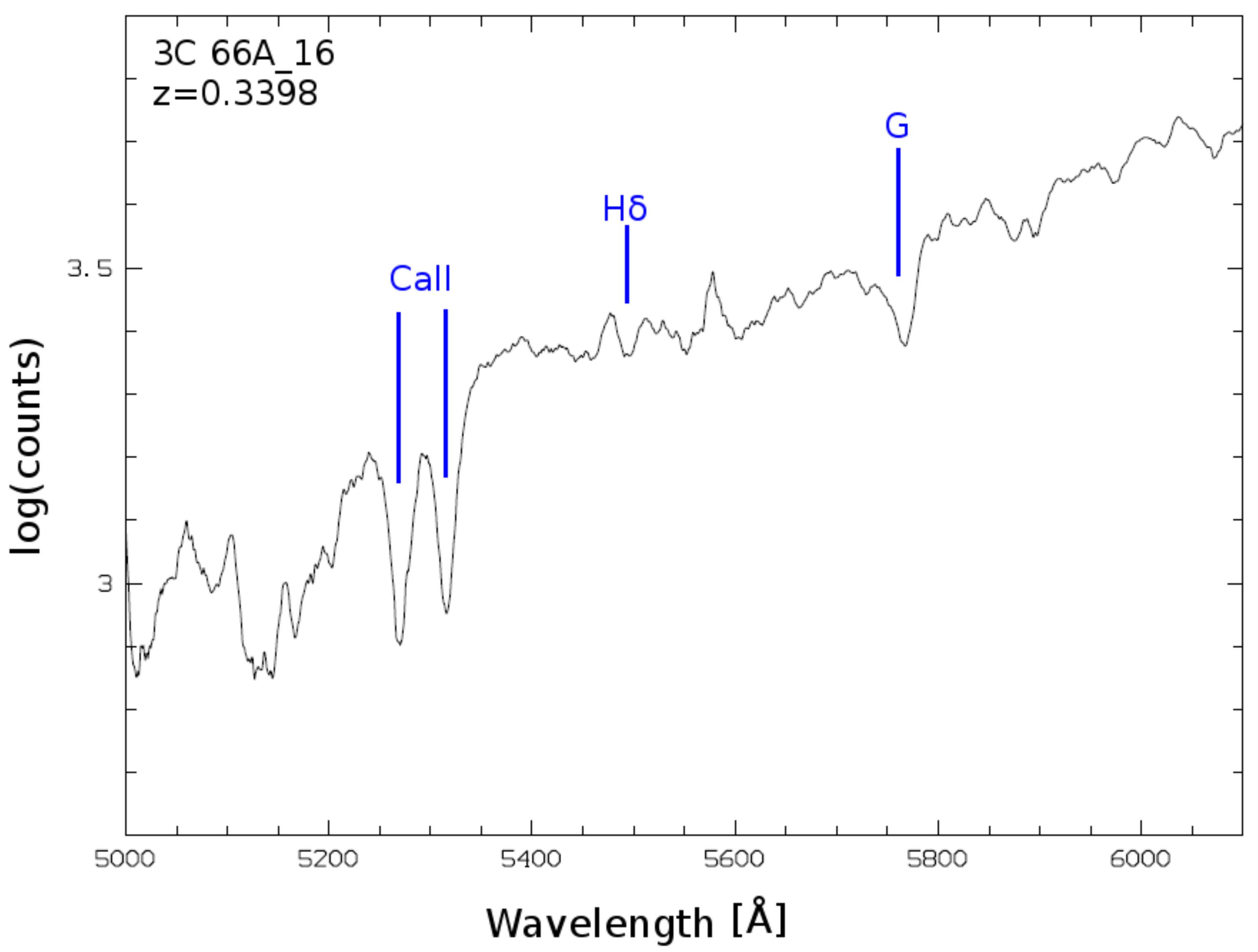}
\caption{Optical spectra ($B600_{-}G5307$ grating) of the six sources
  identified as group members at $z \sim 0.34$ within the field of 3C\,66A,
   in units of counts per {\AA}ngstrom (no
  flux calibration was applied). Spectral features used to determine the
  redshifts of the sources are labelled. A few ``emissions'' are residuals from
  poorly corrected sky lines.}
\label{fig:spectra}
\end{figure*}

The third grouping is comprised by the 6 galaxies with redshifts between $z=
0.3390 \to 0.3402$ ($\langle z \rangle = 0.3400 \pm 0.0006$ rms)
presented in this paper (see Table~\ref{tab:a}). They all have spectral
features typical of early-type galaxies (see Fig.~\ref{fig:spectra}), while
their magnitudes span between $g' \sim 20$ and $22.3$\,mag.  Three of them
are part of the galaxy overdensity identified by \citetalias{BOTT-1976} in
the immediate projected proximity of 3C\,66A (see Fig.~\ref{fig:image}). In
the following subsection we shall further analyse the spatial distribution
and location of these galaxies in the colour-magnitude relation (CMR).

Note that no galaxy was detected at $z\sim 0.444$, which is the usually
quoted redshift for 3C\,66A. This could be, at least partially, the result
of a selection effect, given that an $M^\star$ elliptical galaxy would have
$g' = 22.9$\,mag at that redshift, i.e., too faint for our
spectroscopy. However, we could still have measured spectroscopic redshifts
for E galaxies with $M_{g'}\approx M^\star -1$\,mag at $z =
0.444$. Also note that our data were able to disclose four late-type
  galaxies at $0.43 < z < 0.54$; while none of them is exactly at the
  published redshift for 3C\,66A, at least one of them (object \#10 in
  Table~\ref{tab:a}) would lie within $\sim 3300$\,km\,s$^{-1}$ of the
  blazar, if at that redshift. Thus, our spectroscopic data cannot either
  confirm nor rule out the presence of a cluster at $z=0.444$ in the
  observed field.

\subsection{The reference CRS as a clusters detector} 
\label{sec:CRS}

Taking advantage of the evidence that early type galaxies
  in rich clusters follow a linear colour-magnitude relation (CMR),
  called Cluster Red Sequence \citep[CRS; e.g.:][]{2000AJ....120.2148G},
  and considering the universal properties that the CRS presents: a
  low colour dispersion ($\sigma_{B-R}\sim0.1$\,mag; $\sigma_{g'-i'}\sim0.04$\,mag), a slope that does not change
  significantly with $z$, and a shift to redder colours
  as one progresses in redshift \citep{2004ApJ...614..679L}, the CRS
  of a known cluster, or a reference CRS, can serve as a powerful
  observational tool for detecting galaxy clusters.
 % as well as can to
 %  identify additional galaxies sharing the same colour-magnitude
 %  relation as the cluster red sequence, thus providing new candidate
 %  members that strengthen the detection and characterisation of these
 %  structures.

We use the red sequence of the Virgo cluster, conveniently shifted, as a
  reference CRS to confirm if the spectroscopically identified galaxies are
  members of galaxy clusters. At the same time, this allows us to identify
  new (photometric) candidate members for each cluster. The projected
  spatial distribution of both, confirmed and candidate members, may then
  provide additional support to a real connection between any of the
  clusters and the blazar.
%look for additional galaxies sharing the same colour-magnitude
%relation as the spectroscopically identified groups, thus providing
%new candidate members that would strengthen the detection and
%characterisation of those groups.
%

The adopted CRS was obtained from the photometric analysis for 100 galaxies
of the Virgo Cluster core performed by \citet{2010ApJS..191....1C},
from SDSS (\textit{Sloan Digital Sky Survey}) images in the $ugriz$
bands.  These galaxies allowed us to determine the best-fitting Virgo
CRS, which is represented by:
\begin{equation}
 g'-i' = -0.0482\, g' + 1.6934 .
\label{eq:1}
\end{equation}

We used this CRS shifted to $z =0.34$ and $z =0.02$ (Fig.~\ref{fig:hz}).
Shifts in $g'-i'$ colour and $g'$ magnitude were established using
$K$- and evolutionary corrections computed through the stellar
population synthesis template galaxy models of
\citet{2005MNRAS.361..725B}.
%
%The shifts in colour and magnitude that 
We obtain for the Virgo red-sequence at $z = 0.34$: %are
 $\Delta(g'-i')=0.894$\,mag and $\Delta g'=10.33$\,mag. %, respectively.
 Luminosity distance was computed with the aid of Ned Wright's
 \emph{Cosmology Calculator}
\footnote{\url{http://www.astro.ucla.edu/~wright/CosmoCalc.html}}.

\begin{figure}
\centering
\includegraphics[width=\columnwidth,height=6.3cm]{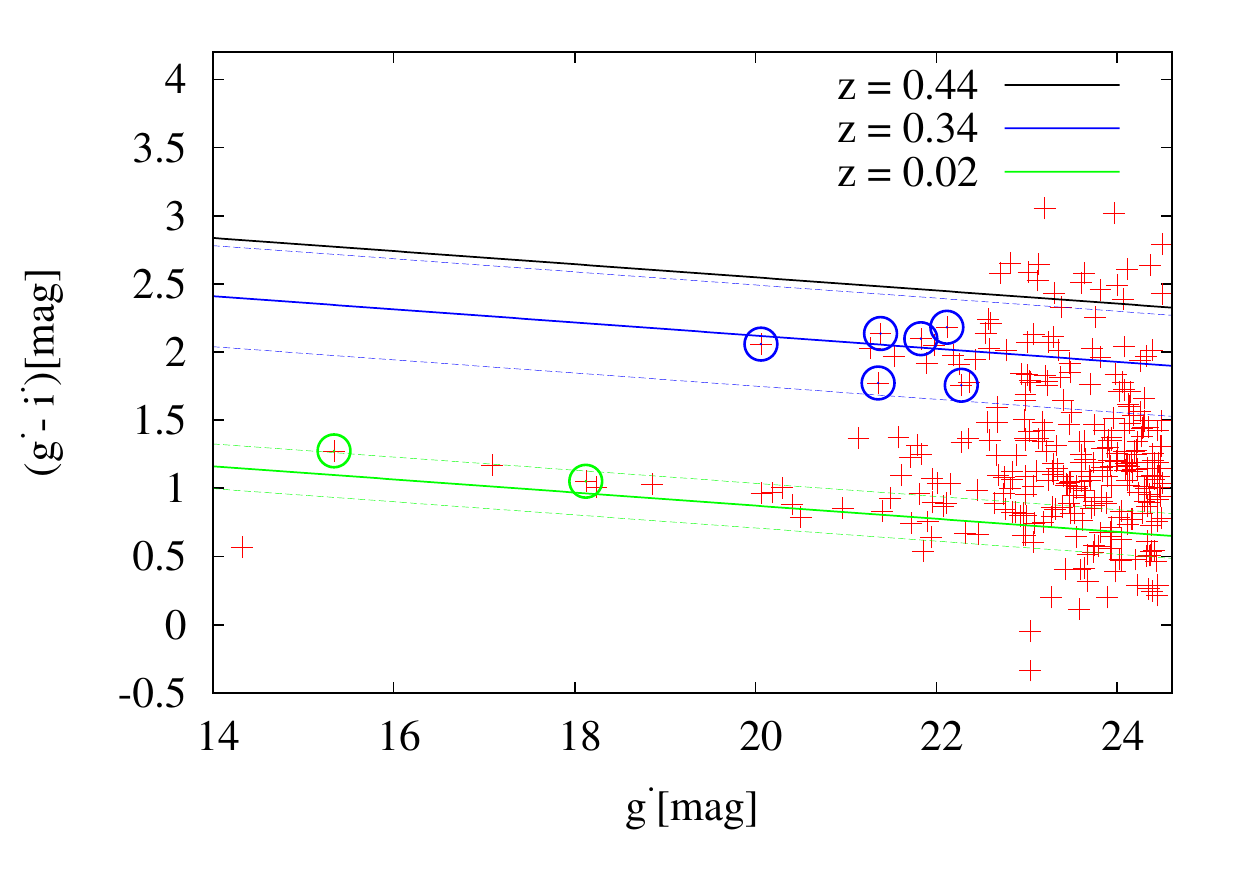}
\caption{Colour-magnitude diagram (CMD) for detected galaxies in the
  field of 3C\,66A (red crosses). Framed crosses within blue circles
  indicate the position of cluster members confirmed through
  spectroscopic redshifts ($z = 0.340 \pm 0.001$).  The blue solid line
  shows the Virgo cluster CRS shifted to
  $z=0.340$. Framed crosses within green circles indicate galaxies with
  spectroscopic redshifts $z \simeq 0.02$. The green solid line
  shows the Virgo cluster CRS shifted to $z=0.02$. Dashed lines show the adopted colour scatter in each
  case. The black solid line shows the Virgo
    cluster CRS shifted to $z=0.444$.}
\label{fig:CMD-z}
\end{figure}

 We plot in Fig.~\ref{fig:CMD-z} the colour-magnitude diagram for all
 extended sources detected in the GMOS field of 3C\,66A, showing also
 the Virgo red-sequence shifted to $z=0.34$ (blue solid line).
The CRS clearly coincides with the position of the
  spectroscopically identified galaxies, confirming the presence of a
  cluster at $z \simeq 0.34$.
To identify other candidate members of this cluster, we
  calculated a maximum colour scatter around the reference CRS and a
  magnitude limit to minimize effects from photometric errors and
  background contamination. The colour scatter was determined based on
  the 6 spectroscopically identified members, $\sigma_{( g' -
    i')}=0.185$\,mag. Following the recommendation of
\cite{2004ApJ...614..679L}, the magnitude cut can be set at 2.0\,mag below
the apparent magnitude corresponding to the break of the typical cluster (Schechter) luminosity
function, $M^*_{R} = -20.3 + 5\log h$, at the cluster redshift. Given
the adopted Cosmology, $M^*_{R} = -21.10$\,mag. This magnitude was
transformed into the Gunn-Sloan photometric system through
  the colours given by \cite{2005MNRAS.361..725B} for galaxies of
  different ages, resulting in $M^*_{g'} = -20.26$\,mag. Therefore, an
  appropriate magnitude cut can be set at $M_{g'} = -18.26$\,mag,
  corresponding to $g' = 23.88$\,mag for this cluster ($z=0.34$).  This allowed
  us to consider as probable members those objects with $g' <
  23.88$\,mag, that are within the $2\,\sigma_{( g' - i')}$ range of
  the CRS (blue dashed lines in Fig.~\ref{fig:CMD-z}). A
    total of 41 candidate members were identified in this way
    (including the 6 confirmed members), and we list them in
    Table~\ref{tab:b}. Galaxy 1456 (\#18 in Table~\ref{tab:a}), however, is
    a background ($z=0.493$) spiral whose colour is redshifted onto the
    $z=0.34$ CRS. We consistently keep it in Table~\ref{tab:b}; it is clear
    that a certain level of contamination still remains within our candidate list.

  Cluster membership probability for each galaxy was determined
    based on their magnitudes, following:
\begin{equation}
 P(g') = \frac{N(g')+ N(g' < m < 23.88)}{N_{T}}.
\label{eq:2}
\end{equation}
Here, $N(g')$ is the number of galaxies with apparent $g'$
  magnitudes within a given range, $N(g' < m < 23.88)$ is the
  number of galaxies with apparent magnitudes between the magnitude cut
  and $g'$, while $N_\mathrm{T}$ is the total number of galaxies in the selected
  range.
  Besides membership probabilities, projected distances to the
    blazar (assuming they are at $z=0.340$) were obtained for the 41 candidate
    members (see Table~\ref{tab:b}). Note that $\approx 50\%$ of the
    candidates would lie within 500\,kpc (projected distance) from the blazar.
\begin{table*}
 \caption{Membership probability of the candidate members of the
   cluster detected at $z = 0.340$: col.\ 1: identification label (ID) by \textsc{SExtractor};
   cols.\ 2 and 3: RA and Dec (J2000); cols.\ 4 and 5:
   $g'$ and $i'$ total integrated magnitudes; col.\ 6: membership probability;
   col.\ 7: projected distance to the blazar, if at $z=0.340$ (the adopted Cosmology gives a scale
   of 293.5\,kpc\,arcmin$^{-1}$).}
 \label{tab:b}
 \begin{tabular}{lcccccc}
  \hline
 ~~~~ID  &RA$_\mathrm{J2000}$ & Dec$_\mathrm{J2000}$ & $g'$        & $i'$         & $P(g')$     &  D$_{C}$\\
\textsc{SExtractor}  &(hh:mm:ss)     & ($^\circ$:$'$:$''$)  & (mag)  & (mag)  &        &  kpc\\\hline
\phantom{1}872$^{\#8}$  & 02:22:50.4  &  43:02:34    &  20.06   & 18.00    &  1.000        &  593.0       \\
1136                  & 02:22:37.0  &  43:02:03     &  21.26   & 19.23    &  0.975      &  141.6         \\
\phantom{1}697$^{\#4}$  & 02:22:51.4  &  43:01:39    &  21.35   & 19.58    &  0.950       &  648.7        \\
1377$^{\#16}$         & 02:22:39.4     &  43:03:04   &  21.38   & 19.25    &  0.926      &   274.1         \\
1456$^{\#18}$         & 02:22:46.9     &  43:03:59   &  21.53   & 19.56    &  0.900 &        669.4       \\ 
1195$^{\#12}$         & 02:22:40.2     &  43:02:20   &  21.83   & 19.73    &  0.878      &  \phantom{2}66.9\\
\phantom{1}717       & 02:22:47.1     &  43:01:22   &  21.89   & 19.97   &  0.853      &  461.0      \\
1142            & 02:22:38.7       &  43:02:13      &  21.97   & 19.92   &  0.829      &  \phantom{2}54.1\\
1243$^{\#15}$         & 02:22:35.1       &  43:02:09  &  22.12   & 19.93   &  0.800        &  241.4        \\
1535               & 02:22:27.9       &  43:02:42   &  22.19    & 20.21   &  0.780       &  649.2         \\
\phantom{1}934    & 02:22:37.6       &  43:01:25    &  22.25    & 20.34   &  0.756      &  236.1          \\
\phantom{1}188$^{\#1}$    & 02:22:35.3  &  42:59:05  &  22.27    & 20.52   &  0.731      &  924.5          \\
\phantom{1}766           & 02:22:53.0 &  43:02:03   &  22.36    & 20.58   &  0.700      &  719.1          \\
1006            & 02:22:39.8           &  43:01:58  &  22.43    & 20.48   &  0.682      &  \phantom{2}50.1\\
1474            & 02:22:48.7           &  43:04:06  &  22.54    & 20.40   &  0.658      &  755.8    \\
1082            & 02:22:40.7           &  43:02:07  &  22.57    & 20.33   &  0.634      &  \phantom{2}59.2\\
1155            & 02:22:48.0           &  43:02:56  &  22.58    & 20.56   &  0.600      &  508.0     \\
\phantom{1}858  & 02:22:38.0           &  43:01:15  &  22.60    & 20.38   &  0.585      &  273.1      \\
\phantom{1}804  & 02:22:45.2           &  43:01:42  &  22.77    & 20.75   &  0.560      &  326.2     \\
1372            & 02:22:38.1           &  43:02:50  &  22.94    & 21.10   &  0.536      &  220.6     \\
2579            & 02:22:38.5           &  43:04:04  &  22.98    & 21.33   &  0.512      &  570.5     \\
\phantom{1}932  & 02:22:44.1           &  43:02:01  &  22.98    & 21.29   &  0.487      &  243.8     \\
2081            & 02:22:38.3           &  43:04:18  &  23.00    & 20.93   &  0.463      &  639.7     \\
\phantom{1}772  & 02:22:38.1           &  43:00:53  &  23.01    & 21.17   &  0.439      &  375.7     \\
1498            & 02:22:42.9           &  43:03:41  &  23.03    & 21.24   &  0.414      &  488.2     \\
1388            & 02:22:36.1           &  43:02:44  &  23.04    & 21.26   &  0.390      &  257.4     \\
1511            & 02:22:39.4           &  43:03:42  &  23.07    & 20.94   &  0.365      &  460.0     \\
1111            & 02:22:49.1           &  43:02:52  &  23.20    & 21.38   &  0.341      &  553.1     \\
2510            & 02:22:37.7           &  43:04:29  &  23.22    & 21.46   &  0.317      &  697.2     \\
1910            & 02:22:23.7           &  43:03:23  &  23.22    & 21.43   &  0.292      &  928.2     \\          
\phantom{1}198  & 02:22:46.4           &  42:59:59  &  23.24    & 21.16   &  0.268      &  729.0     \\          
1577            & 02:22:39.2           &  43:03:46  &  23.29    & 21.18   &  0.243      &  479.9     \\          
\phantom{1}532  & 02:22:34.7           &  42:59:40  &  23.35    & 21.33   &  0.219      &  770.3     \\          
\phantom{1}689  & 02:22:52.0           &  43:01:35  &  23.37    & 21.55   &  0.195      &  684.4     \\          
\phantom{1}456  & 02:22:54.0           &  43:00:57  &  23.40    & 21.75   &  0.170      &  847.0     \\          
2478            & 02:22:44.6           &  43:04:50  &  23.47    & 21.55   &  0.146      &  836.6     \\          
1173            & 02:22:41.8           &  43:02:21  &  23.48    & 21.63   &  0.121      &  134.0     \\          
1936            & 02:22:27.6           &  43:03:39  &  23.69    & 21.93   &  0.097      &  782.4     \\           
2410            & 02:22:41.3           &  43:04:41  &  23.72    & 21.69   &  0.073      &  754.0     \\         
\phantom{1}723  & 02:22:46.7           &  43:01:20  &  23.75    & 21.49   &  0.048      &  447.5     \\           
1553            & 02:22:28.3           &  43:02:42  &  23.81    & 21.85   &  0.024      &  628.4     \\                                                                                                       
% 901           & 4692.717             &  1787.417            &  23.9489             & 22.2109      & 0.047      &  586.621          \\           
%1130           & 2835.698             &  2277.348            &  23.9811             & 22.1448      & 0.023      &  101.676          \\ 
\hline
\multicolumn{7}{l}{\#1, \#4, \#8, \#12, \#15, \#16 and \#18: object numbers in Table~\ref{tab:a}.}\\
\end{tabular}
\end{table*}

The spatial distribution of candidate members can be further illustrated
by means of a density map \citep{GdCO2000}. Fig.~\ref{fig:image}
shows contours of projected galaxy density, using a fixed Gaussian kernel with
$\sigma = 1$\,arcmin, where candidates are weighted by $P(g')$, as given in
Table \ref{tab:b}. The projected distribution of $z=0.340$ candidates is
strongly concentrated to 3C\,66A, with the maximum density centred very
close to the
blazar. The central peak was found to be significant at a $>3
\sigma$ level, through a significance map constructed from 500 synthetic
data sets \citep[see][for details on significance maps computation]{GdCO2000}.

The integrated luminosity of the 41 candidate member galaxies
(excluding the blazar's host) amounts to $\mathcal{L}_\mathrm{tot}(g')
= 3.4 \times 10^{11}\,\mathcal{L}_\odot$, giving an upper
  limit of total stellar mass $\mathcal{M}_\star \approx 5 \times
10^{12} \, \mathcal{M}_\odot$ (without considering any weight
  for the galaxies), consistent with a poor cluster of galaxies
\citep[see e.g.][and references therein]{BCSZ12}, as proposed by
\citet{WESY-1993}.

As discussed in Section~\ref{sec:intro}, and considering that
BL\,Lac objects are usually hosted by giant elliptical galaxies, 
typically associated with groups or galaxy clusters, we propose that the
host galaxy of 3C\,66A is a member of the cluster at $z=0.340$ found in this
work.
Assuming that the host galaxy is typical for TeV emitting BL\,Lacs
(see section~\ref{sec:thg}), its apparent magnitude should be $g' \sim
20.5$\,mag, i.e., slightly fainter than the brightest galaxy in the
group (slit \#8). The cluster, then, seems to be dominated by these two
bright ellipticals. 

On the other hand, another red sequence, highlighted by some of the
brightest objects in the field, can be visually identified in
Fig.~\ref{fig:CMD-z}. We found that it can be fitted by shifting the Virgo
CRS to $z=0.02$, and the two galaxies with (spectroscopic) $z=0.02$ lie on
this red sequence; this allows us to verify the presence, within the
analysed field, of several members of the foreground group associated with
Abell\,347. Note that the typical Abell radius for a galaxy cluster
  is $1.7$\,arcmin\,$z^{-1}$, so a cluster at $z = 0.02$ would cover a diameter of
  $170$\,arcmin. This indicates that the GMOS field of view ($5.5 \times
  5.5$\, arcmin$^{2}$) allows us to observe only a small portion of the
  cluster (most probably at its outskirts).
Since in this case we are sampling the faintest portion of the red sequence
(where colour scatter is larger), in order to identify candidate members of
this group we adopted a $(g' - i')=0.16$\,mag colour range (green dashed
lines in Fig.~\ref{fig:CMD-z}), and a magnitude cutoff at $g' = 20.2$\,mag,
which corresponds to $M_{g'} = -14.5$ at $z = 0.02$, i.e. within the
early-type dwarf galaxies domain, where the CMR is still well defined for
nearby clusters \citep[see][]{2008MNRAS.386.2311S, CBC-2015}.
In this way, besides the 2 spectroscopically confirmed galaxies, \#14
  and \#22, we identify \textbf{5} additional candidate members. Of these,
  galaxy G2 in \citetalias{1997MNRAS.284..599B} ($z=0.0677$, i.e. a
  neighbour of the radio source 3C\,66) and object \#3 ($z=0.1521$) are
  background interlopers in the early-type galaxies CMR at $z=0.02$.
The other candidates (including objects \#5 and \#11, for which we could
obtain no spectroscopic redshift, due to their very low surface
brightnesses), however, show clear dE (dwarf elliptical) morphologies; this
has been shown to be an excellent membership indicator for low-redshift
clusters \citep[e.g.][and references therein]{CB05}, so we retain them as
probable members of the $z=0.02$ group.

Note that no CRS for $z \sim 0.0677$ is evident in Fig.~\ref{fig:CMD-z}; the
three galaxies at this redshift in \citetalias{1997MNRAS.284..599B} are
spirals, so they may belong to a loose foreground grouping of late-type galaxies.
We also show in Fig.~\ref{fig:CMD-z} the Virgo CRS shifted to $z =
0.444$. No red sequence is observed at this $z$ value down to $\sim M^\star$
($g'=22.9$\,mag, see Sect.~\ref{sec:idgalg}).
%  Due to the faintness of the
% objects close to this linear relationship, we can not
% establish its morphology in the image and neither to confirm a
% possible cluster at $z = 0.44$.

\subsection{Analysis of the optical spectrum of 3C\,66A}  
\label{sec:spec3C66A}

Due to its inherent nature of non-thermal emission, a proper analysis of
any blazar spectrum is a difficult task. In our specific case, as we
will see in the next section, the underlying galaxy hosting 3C\,66A
(most likely a gas-poor early-type system) seems to be too faint to
leave any explicit signature in the target integrated spectrum. The
resulting spectral energy distribution (SED) is, therefore, expected to
lack any striking absorption feature, and actually this enhanced
``smoothness'' is a recognised drawback when trying to constrain the
blazar distance (and therefrom its related physical properties)
through a redshift measurement. 

\begin{figure}
\centering
\includegraphics[width=\columnwidth]{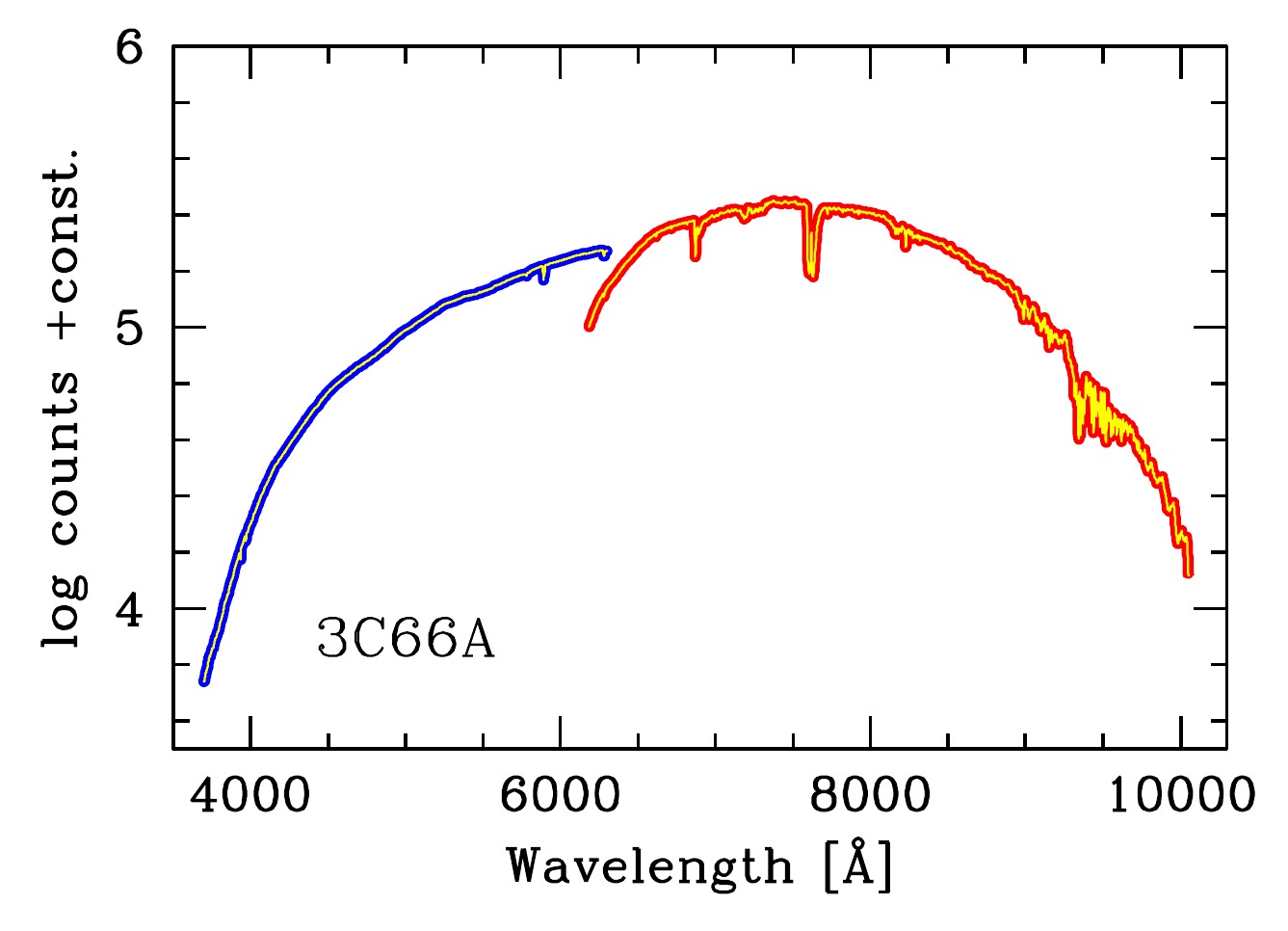}
\caption{Wavelength branches of 3C\,66A's spectrum. The branch
    obtained with the $B600_{-}G5307$ grating (blue line) covers a
    spectral range between 3500\,\AA\ and 6500\,\AA, while the branch
    obtained with the $R400_{-}G5325$ grating (red line) covers a spectral
    range between 6100\,\AA\ and 10\,000\,\AA.}
\label{fig:blazar_fig_all}
\end{figure}

A hint pointing to a quite ``blue'' object does emerge from our
photometry, once considering the likely effect of redshift on target
apparent colour (namely, $g'-i' = 0.91$).  The two wavelength branches
of 3C\,66A's spectrum, respectively for the $B600_{-}G5307$ and
$R400_{-}G5325$ gratings are displayed in Fig.~\ref{fig:blazar_fig_all}.

\begin{figure}
\centering
\includegraphics[width=\columnwidth]{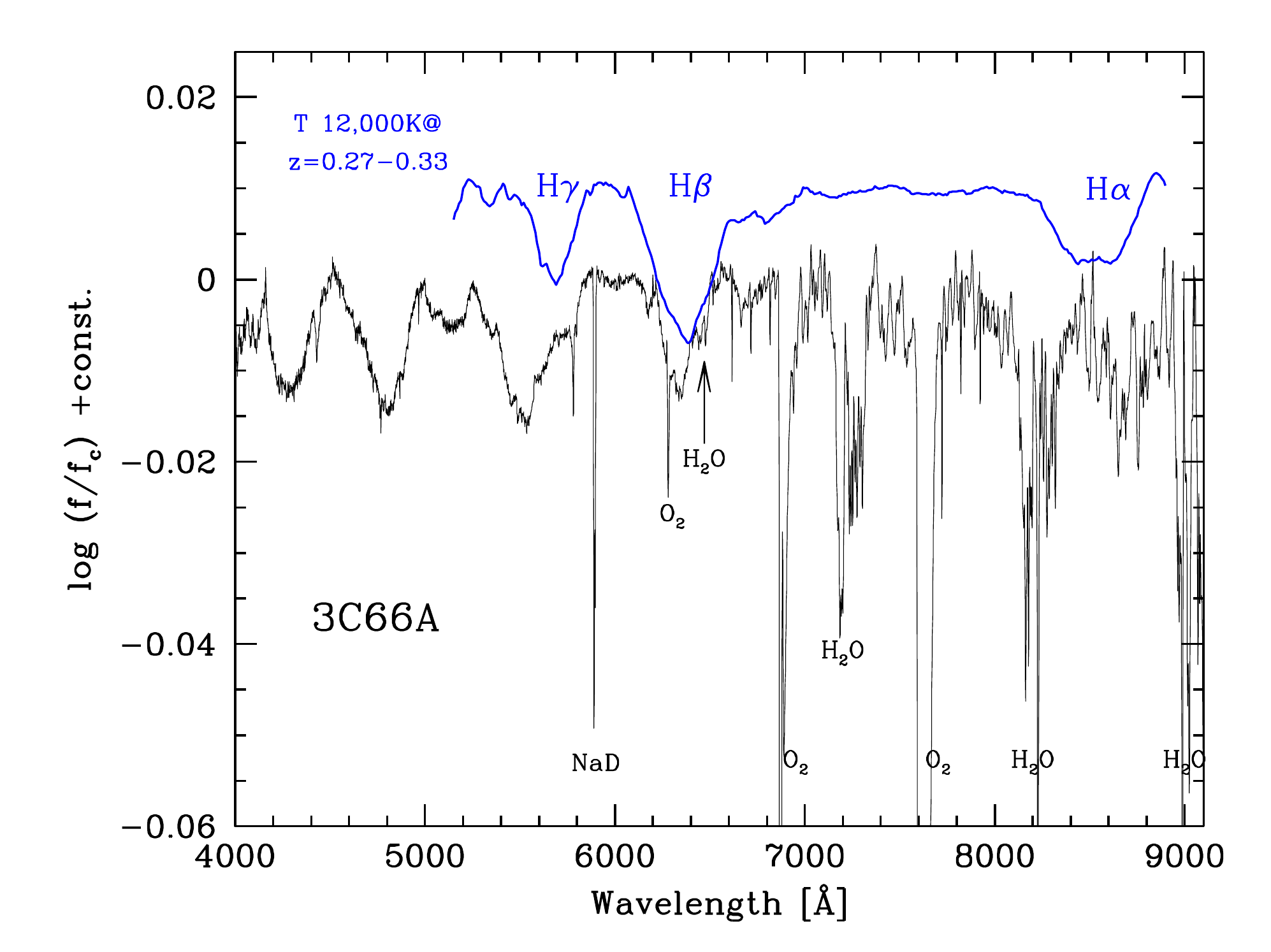}
\caption{The linearised 3C\,66A spectrum, compared with a \textsc{Bluered}
synthetic spectrum of a stellar model atmosphere with $T=12\,000$\,K
shifted within the range $0.27<z<0.33$ (blue line).
The Balmer sequence of the H$\gamma$, H$\beta$ and H$\alpha$
absorption lines might be recognised to match the blazar spectrum.
See text for a discussion.}
\label{fig:spec-blaz-norm}
\end{figure}

At first glance, one sees that a more thorough analysis could be
enabled if the bell-shaped instrumental response is removed, for
instance by normalising the observed spectrum with respect to its
pseudo-continuum. This task has been carried out with the appropriate
\textsc{midas} routine (i.e. \textsc{normalise/spectrum} in the
\textsc{Longslit} context), by interactively identifying the upper
envelope in the logarithmic count domain to be eventually subtracted
to the observations.  The blue and red branches of the spectrum have
then been carefully matched around 6100 \AA, and a unique spectrum,
between 4000 and 10\,000 \AA, has been obtained
(Fig.~\ref{fig:spec-blaz-norm}).

After linearisation, a more entangled plot begins to appear from our
data, clearly putting in evidence a number of thin but significant
absorption features. To a closer analysis, one has first to report the
pervasive contribution of telluric absorptions. Both O$_2$ and water
vapour (H$_2$O) bands strongly affect the target emission and severely
tackle our chance to single out any genuine target feature longward of
$\sim 7000$ \AA.  The telluric pattern, however, fairly well compares
with the study of \citet{1994MNRAS.267..904S} and these features
clearly stand out, in our spectrum, for their sharp profile.

Quite more importantly, on the contrary, a number of broadened
absorption features also mark the linearised spectrum, with prominent
bands easily recognised about 4300, 4800, 5550, 6350 and 8700\,\AA\ (see, again, Fig.~\ref{fig:spec-blaz-norm}).  
Such broadened
features are certainly of extragalactic origin, and might be
suggestive of some interloping gas clouds, which selectively absorb
the 3C\,66A luminosity along the line of sight. No obvious constrain
can be posed to this scenario, of course, but a speculative fit with a
redshifted synthetic spectrum from the {\sc Bluered} stellar library
\citep{BBC2008} for a $T = 12\,000$\,K (metal-poor) model
atmosphere might lead to identify some of the observed bands with the
H$\alpha$, H$\beta$ and H$\gamma$ Balmer lines, as provided by
foreground gas clouds up to $z\la 0.33$, as sketched in the figure.
If this is the real case, then a lower limit can be placed to 3C\,66A
redshift at
$z_{3C66A} \ga 0.33$.

\subsection{The host galaxy}
\label{sec:thg}

In this section we briefly analyse the image of the blazar, in order to see
if we can detect its host galaxy, to further constrain or confirm the
blazar's redshift.  We used \textsc{iraf} task
\textsc{psfmeasure} to fit a Moffat function to the PSF of the blazar and 12
stars on the central chip of the GMOS array (the same where the blazar
image is). This was done on the $g'$ image only, because in the $i'$ image the AGN
is slightly saturated. After correcting for PSF trends with position on the
chip, we find that the FWHM of the blazar's PSF is marginally larger ($2.6
\,\sigma$) than the mean value corresponding to stars. For the Moffat $\beta$
parameter, in turn, the blazar's value is $4.0 \,\sigma$ larger than for the
stars. This would mean that the host galaxy is marginally resolved in our
images, in agreement with \citet{1996ApJS..103..109W}. However, the presence
of haloes of stray light around the AGN and bright stars in our images makes
this result uncertain.

We then simulated images of an elliptical galaxy (represented by a de
Vaucouleurs model) with $M_R =-22.5$\,mag and $R_\mathrm{eff}=10$\,kpc
\citep{SRC-2013, STF05} plus a Moffat model representing the AGN, at
different redshifts. Galaxy models 1\,mag fainter/brighter than the mean
adopted absolute magnitude were also considered, and their apparent
magnitudes were calculated using $K$- and evolutionary corrections
following \citet{2005MNRAS.361..725B}.  Due to the mentioned stray
light halo around the blazar's image, all we can say is that detection
of a typical host galaxy at $z=0.340$ would have been marginal, at
best. In any case, it is clear that we can definitely rule out a low
redshift value such as $z=0.067$, corresponding to the foreground
grouping of galaxies mentioned in
section~\ref{sec:idgalg}.

\section{SUMMARY AND CONCLUSIONS}
\label{sec:syc}

We have spectro-photometrically analysed the close environment of 3C\,66A
using two-band ($g',\, i'$) optical images of a $5.5 \times
5.5$\,arcmin$^{2}$ field centred on the blazar, along with multi-object
spectroscopy obtained with the Gemini North telescope and the Gemini Multi
Object Spectrograph (GMOS). We obtained spectra for the blazar as well as
twenty-four objects in the field, and we were able to measure reliable
redshifts for 15 galaxies in the blazar's line of sight, spanning the range
$0.02<z<0.53$.  We found no evidence confirming the published
  redshift ($z=0.444$) for 3C\,66A, although we cannot formally rule it out
  either.  Instead, we could firmly establish the presence of a galaxy
group at $\langle z \rangle = 0.3400 \pm 0.0006$ in the close environment of
3C\,66A. In particular:

\begin{itemize}
\item We identified two concentrations of galaxies (in redshift space) along
  the blazar line of sight, at $z\approx 0.020$ and $z \approx 0.340$,
  respectively.  The first of them corresponds to the poor cluster WBL\,069,
  associated to Abell\,347 \citep{WBB1999}; we identified two new members of
  this cluster.  The second one is composed of six objects with redshifts
  between $z=0.3390$ and $z=0.3402$; we classify almost all of them as early
  type galaxies.  On the other hand, there is a galaxy in our GMOS field
  that seems to belong to another loose grouping in the line of sight at $z
  \approx 0.067$, associated with the radio-source 3C\,66.

\item Using the Virgo Cluster red sequence \citep[from data published
  in][]{2010ApJS..191....1C} as a reference CRS, shifted to $z=0.02$ and
  $z=0.34$, we were able to identify two structures (i.e., ``red
  sequences'') traced by early-type galaxies on the %$g'$ vs. $g'-i'$
  colour-magnitude diagram. The six galaxies with $\langle z \rangle=0.340$
  and the two galaxies with $\langle z \rangle =0.020$, coincide with the
  position of the Virgo CRS shifted to these same $z$ values, respectively,
  thus supporting the presence of two galaxy clusters, one at $z=0.340$ and
  other at $z=0.020$.
%lie on the redder and bluer of these sequences, respectively.
%We established the Virgo's CRS using data published in 
%\citet{2010ApJS..191....1C} and we shifted it to $z=0.02$ and $z=0.34$, 
%finding a very good match with both red sequences, which
%further strengthens the evidence for the presence of two galaxy
%concentrations at those redshifts.

\item Adopting a colour dispersion $\sigma_{( g'- i')}=0.04$\,mag for the
  cluster detected at $z=0.02$, we identified $7$ candidate members
  (including $2$ spectroscopically confirmed members and 2 confirmed
    background objects).  On the other hand, for the $z=0.34$ cluster, the
  colour dispersion (as measured from the 6 spectroscopically confirmed
  members) is $\sigma_{( g' - i')}=0.185$\,mag. This allowed us to identify
  41 candidate members from the CMD (including $6$ spectroscopically
  confirmed members and one confirmed background galaxy). The
     projected spatial distribution of $z=0.340$ candidates shows a significant
    concentration centred on 3C\,66A's position.

\item It was not possible to detect the 3C\,66A host galaxy on the
  images. Simulations of a de Vaucouleurs profile for the host galaxy
  ($M_R=-22.5$\,mag and $R_\mathrm{eff}=10$\,kpc, \citealp{SRC-2013,
    STF05}) combined with a Moffat function for the AGN seeing
  convolved PSF, showed that the host galaxy would have been, at best,
  marginally detected if at $z = 0.34$.  At the same time, membership
  of 3C\,66A to any of the lower redshift groups at $z = 0.067$ and $z
  = 0.020$ can be ruled out.

\item Analysing the blazar's normalised optical spectrum between
  $4000$\,\AA--$10\,000$\,\AA, we were able to relate some absorption
  features with contributions of telluric origin, while shallow and
  broad absorption bands around $4300$, $4800$, $5550$, $6350$, and
  $8700$\,\AA\ could be associated with the foreground
  intergalactic medium (IGM). By fitting a stellar synthetic spectrum
  ($T=12\,000$~K) at varying redshifts, it was possible to tentatively
  identify those absorption bands with the Balmer series (H$\alpha$,
  H$\beta$ and H$\gamma$) produced in the IGM at $z\la 0.33$. In this way, a
  redshift lower limit $z_{3C66A} \ga 0.33$ can be established for
  3C\,66A. This is consistent with the firm lower limit $z\ge 0.3347$ and
  the 99.9\% confidence upper-limit $z<0.41$ given by
  \cite{2013ApJ...766...35F}.

  Finally, considering these $z$ lower/upper limits for 3C\,66A, and
  taking into account the fact that BL\,Lac hosts are, typically,
  early type galaxies associated to a group or a galaxy cluster
  \citep{2000ApJ...532..816U}, we propose that the host galaxy of
  3C\,66A belongs to a galaxy cluster at $z = 0.340$.
 
\end{itemize}

\section*{Acknowledgements}
Based on observations obtained at the Gemini Observatory (program
GN-2009B-Q-2), which is operated by the Association of Universities for
Research in Astronomy, Inc., under a cooperative agreement with the NSF on
behalf of the Gemini partnership: the National Science Foundation (United
States), the National Research Council (Canada), CONICYT (Chile), Ministerio
de Ciencia, Tecnolog\'{\i}a e Innovaci\'on Productiva (Argentina), and
Minist\'erio da Ci\^encia, Tecnologia e Inova\c{c}\~ao (Brazil). Data were
acquired through the Gemini Science Archive and processed using the Gemini
IRAF package.

This work has been partially supported with grants from La Plata National
University (Argentina) and The Argentinian National Gemini Office.  We thank
Dr. Germ\'an Gimeno, Gemini staff member, for his advice in the process of
Nod\&Shuffle data reduction, and Dr. Anal\'{\i}a Smith Castelli for her
recommendations on data analysis.

The authors want to thank the anonymous referee for valuable
  comments that really helped to improve this paper.

%%%%%%%%%%%%%%%%%%%%%%%%%%%%%%%%%%%%%%%%%%%%%%%%%%



\bibliographystyle{mnras}
\bibliography{3C66Av1p15} % if your bibtex file is called example.bib

%%%%%%%%%%%%%%%%%%%%%%%%%%%%%%%%%%%%%%%%%%%%%%%%%%

%%%%%%%%%%%%%%%%% APPENDICES %%%%%%%%%%%%%%%%%%%%%

%%%%%%%%%%%%%%%%%%%%%%%%%%%%%%%%%%%%%%%%%%%%%%%%%%

%\appendix

%\section{CANDIDATE DWARF GALAXIES IN THE FOREGROUND GROUP ($z=0.02$)}
%\label{sec:dwarfs}

% Don't change these lines
\bsp	% typesetting comment
\label{lastpage}
\end{document}